\documentclass[%
 reprint,
 amsmath,amssymb,
 aps,
notitlepage,
onecolumn ,
]{revtex4-2}

\usepackage{graphicx}
\usepackage{dcolumn}
\usepackage{bm}
\usepackage{physics} 
\usepackage{braket}
\usepackage{mathtools}
\DeclarePairedDelimiter{\ceil}{\lceil}{\rceil}
\usepackage{hyperref}
\usepackage{xcolor}
\definecolor{Navy}{RGB}{0, 0, 128}

\begin{document}

\preprint{}

\title{Trap Frequency Measurement with a Pulsed Atom Laser}

\address{Department of Quantum Science and Technology, Research School of Physics, The Australian National University, Canberra, ACT 2601, Australia
}
\email{bryce.henson@anu.edu.au}

\address{Department of Quantum Science and Technology, Research School of Physics, The Australian National University, Canberra, ACT 2601, Australia}

\author{B. M. Henson}
\email{bryce.henson@anu.edu.au}
\author{K. F. Thomas}
\author{Z. Mehdi}
\author{T. G. Burnett}
\author{J. A. Ross}
\author{S. S. Hodgman}
\author{A. G. Truscott}
\email{andrew.truscott@anu.edu.au}
\affiliation{Department of Quantum Science and Technology, Research School of Physics, The Australian National University, Canberra, ACT 2601, Australia}

\date{\today}

\begin{abstract}
We describe a novel method of single-shot trap frequency measurement for a confined Bose-Einstein Condensate, which uses an atom laser to repeatedly sample the mean velocity of trap oscillations as a function of time. 
The method is able to determine the trap frequency to an accuracy of 39~ppm (16~mHz) in a single experimental realization, improving on the literature by a factor of three.
Further, we show that by employing a reconstructive aliasing approach our method can be applied to trap frequencies 
more than a factor of 3 greater than the sampling frequency.
\end{abstract}

\maketitle


\section{Introduction}

Approximating a confining potential near its minima as harmonic is ubiquitous throughout the field of physics.
Under this approximation the potential is completely characterized by the mass and the trap frequency which, in the absence of damping, corresponds to the spatial oscillation frequency. 
In cold-atom experiments the vast majority of confining potentials formed by magnetic or optical fields, are well approximated as harmonic. 
The trap frequency is therefore a key characterization parameter for describing properties of these systems, such the condensation temperature or the density distribution \cite{Pethick2008}.

Measurements of trap frequency can be used for curvature sensing, providing a means to measure the second spatial derivative of a perturbing potential. This is particularly well suited to the sensing of weak perturbing potentials that are spatially localized \cite{Harberphd}. In this way, trap frequency based sensors are complementary to interferometric sensors that typically measure the first derivative
and curvature \cite{PhysRevA.99.033619, PhysRevLett.118.183602} of a potential, such as gravity, over a wide spatial extent .
Trap frequency measurement has enabled precision sensing methodologies across a wide range of fields such as a single electron sensitive electrometer \cite{Cleland1998}, measurement of the proton magnetic moment \cite{Schneider1081}, and measurement of the Casimir-Polder force with a Bose Einstein condensate (BEC) \cite{PhysRevA.72.033610,PhysRevLett.98.063201,Harberphd}.

There has been little development of techniques for trap frequency measurement in BEC experiments over the last decade, despite being a routine part of characterization and having sensing applications.
Furthermore, 
the uncertainty of established measurement techniques due fundamental effects such as shot noise and the density distribution of a BEC in position/momentum has not been well studied.
Improving the precision, speed and convenience of trap frequency measurement will allow higher precision investigations of BEC physics including many body physics phenomena (e.g. quantum depletion \cite{PhysRevLett.117.235303,rossQD,Tenart2021}, thermalization \cite{Brown2018}, collective excitations \cite{Bartenstein2004}, and phase transitions \cite{PhysRevLett.98.040401,PhysRevA.101.053633}) and improve the capability of curvature sensing.

In recent work, we have found existing trap frequency measurement techniques to be a limiting factor in the measurement of a tune-out frequency in Helium \cite{henson2021precision}. Specifically, the fractional uncertainty in the trap frequency was found to limit the use of the tune-out frequency measurement as a rigorous test of quantum electrodynamics (QED).

In this work, we present a novel method for measuring the trap frequency of a confined BEC using a pulsed atom laser \cite{Manning:10}
which we have employed in a number of recent works
\cite{henson2021precision,rossQD,Henson13216}.  
We demonstrate this technique allows for measurement at a precision beyond existing methods, achieving a fractional uncertainty of 39~ppm in a single experimental run.
We also demonstrate that our implementation of this method approaches the fundamental limiting uncertainty for this technique.

\section{Prior Approaches to Trap Frequency Measurement}
The key metric that we will use to compare sensing methods is the fractional uncertainty in the measured trap frequency in a single experimental run (termed a \textit{shot}), which consists of the creation of a BEC and the measurement protocol. 
For methods that require multiple shots to form a trap frequency measurement we have assumed inverse square root scaling of the uncertainty with the number of shots. 
As a general remark, methods that require lower number of shots to make a measurement are generally preferable as they are 
able to resolve faster changes in the trap frequency.
There are two main approaches to measuring trap frequencies in cold atoms experiments: excitation with parametric heating and measurement of trap oscillations. 

\textit{Trap driving} is a popular approach  \cite{smirne2005experiments,LHumbertPHD,wangphd,OE.20.003711,MAKHALOV2015327,PhysRevA.88.043406}, and involves the transfer of kinetic energy to the trapped atoms via periodic modulation of the trap potential. 
The excitation is maximized when the drive frequency is at the fundamental or harmonics of the trap frequency (parametric heating), and is measured as loss of high energy atoms or heating. 
Typically, the measured frequency response in these methods is in general rather wide, leading to a poor uncertainty in the measured trap frequency.
Furthermore, the harmonic approximation of the trapping potential often breaks down in these methods, as atoms must sample increasingly high-energy anharmonic regions of the trapping potential in order to be lost. This results in a complex shape of the measured frequency response, and can introduce systematic errors.
Finding this response also takes many experimental cycles, requiring a long acquisition time. 
To our knowledge, the best fractional uncertainty such a method has been able to achieve is $4\times10^{-3}$ (4000~ppm) per square root shots  \cite{wangphd}.

The second class of methods which we call \textit{kick-and-drop} induces a center of mass oscillations in the BEC (\textit{kick}), followed by the BEC being released from the trap (\textit{drop}) after a variable time delay. 
By detecting the mean position (typically with absorption imaging) of the freely evolving BEC, the mean velocity at the time of release can be deduced. 
The trap frequency is then found by fitting these measurements (acquired over many experimental cycles) of the in-trap velocity as a function of delay time \cite{yibophd,Whitlockphd,Harberphd,Altmeyerphd,PhysRevLett.98.040401,berradaphd,MatthiasTheisphd}.
To date the highest precision application of this approach is by Harber \textit{et al.} with a fractional error of $4\times10^{-4}$ (110 ppm) per square root shots \cite{Harberphd}.

Some non-destructive methods, which allow for multiple measurements with the same BEC, have also been used to measure the trap frequency. 
These include Faraday imaging \cite{doi:10.1063/1.4818913}, interferometric approaches \cite{PhysRevA.75.033803}, echo spectroscopy \cite{Oblak2015}, and relative phase shift \cite{Kohnen_2011}.
Faraday imaging has demonstrated the highest precision, of the aforementioned techniques, with a fractional error of $3\times10^{-4}$ (300 ppm) for a single-shot \cite{doi:10.1063/1.4818913}. 
The method we describe below 
combines
the kick-and-drop method with a (effectively) non-destructive probe of the motion of the BEC.

\begin{figure}
\centering\includegraphics[width=\textwidth]{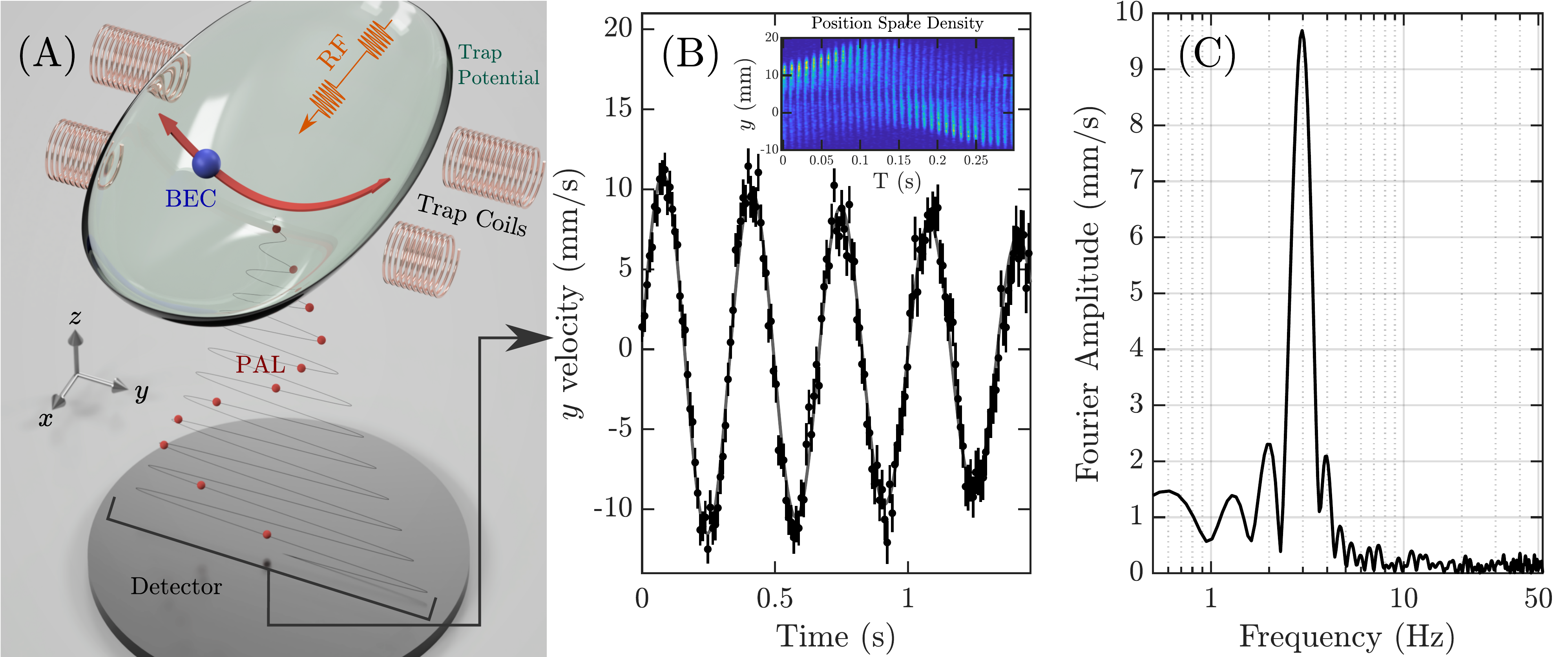}
\caption{
(A) Experimental schematic. The velocity of a oscillating BEC in a magnetic trap is periodically sampled by transferring a small fraction into the untrapped state using pulses of radio frequency (RF), forming a pulsed atom laser (PAL).  The PAL falls under gravity onto a detector below the trap which 
measures
the mean position of each pulse and in turn the velocity of the BEC. Gravity is aligned down the page, anti-parallel to the $z$ direction.
(B) A sine wave fit, of the form of \autoref{model}, to $y$-axis data for a single experimental run, with sampling frequency $f_s=105.2$~Hz. 
Each point represents the center-of-mass of a single atom laser pulse, which has then been converted to near field velocity, with error bars indicating the standard deviation of the density distribution of that pulse. 
We determine an apparent frequency of $f_a = 2.998(6)$~Hz and a damping rate of $\lambda=0.30(4)$~s$^{-1}$. 
(B Inset) Two dimensional histogram showing changes in the density distribution of the PAL during oscillation. Inset shows the integration of 8 shots.
(C) Discrete Fourier transform of the sampling series, shown in (B), which has a peak frequency of $2.996(13)$~Hz, with a signal-to-noise ratio of $86$, and a cutoff frequency equal to the Nyquist frequency $f_N=52.6$~Hz. The true frequency was determined 
by reconstructive aliasing
to be $415.13(1)$~Hz (see \autoref{sec:aliased} and \autoref{fig:aliasing}).
}
\label{fig:fittedoscillations}
\end{figure}

\section{Atom Laser Trap Frequency Measurement}

Our novel method involves setting a BEC in motion and uniformly outcoupling a small fraction of the atom cloud from a magnetic trap by using weak, radio frequency (RF) pulses to transfer some atoms into an untrapped state.
By then measuring the mean position of these pulses in the far field we can reconstruct the in-trap velocity of the BEC as a function of time (as in the \textit{kick-and-drop} method) with only a single experiment.

Our implementation of the method uses a BEC of Helium atoms in the long-lived metastable $2^{3\!}S_1$ excited state (He*) \cite{PhysRevLett.103.053002}. 
Due to the large internal energy of the $2^{3\!}S_1$ state, single atom detection with high spatial and temporal resolution is possible via a micro-channel plate (MCP) and delay line detector (DLD) system \cite{Manning:10}.
After forming the BEC of He* in a bi-planar quadrupole Ioffe configuration (BiQUIC) trap with the methods outlined in \cite{DALL2007255}, we use a short (50~\textmu{}s) pulse of a magnetic field gradient to set the atoms into oscillation, in all three dimensions, around the magnetic trap minimum. 
The trap frequencies for the experiments in this work are  $(f_x,f_y,f_z)\approx (51,412,415)$~Hz.

While oscillating, the velocity of the BEC is periodically sampled by a \textit{pulsed atom laser} (PAL), which weakly out-couples a small fraction of atoms ($\sim1 \% $) from the magnetic trap with pulses of RF radiation resonant with the $m_J=1$ (trapped) $\rightarrow m_J=0$ (untrapped) state Zeeman splitting. 
For atoms at the trap minimum this corresponds to a transition frequency of $\sim1.7$ MHz. 
As the BEC oscillates, it will sample different magnetic field strengths and hence experience different state splitting.
To ensure near uniform outcoupling as the BEC oscillates we use a short ($\sim5$~\textmu{}s) pulse duration which ensures a spectral width ($\sim5$~KHz) wide enough to capture all positions in the oscillation cycle.

The timescale of these pulses is far below the characteristic timescale of the in-trap dynamics, and they can hence be considered as near instantaneous sampling. 
After falling approximately $850$~mm under gravity, the spatial-temporal distribution of the atom laser in the far field is then measured by the MCP-DLD system, as shown in \autoref{fig:fittedoscillations}~A. 
From this we use the mean position of the detected atoms in the pulse to reconstruct the mean velocity of the condensate in the near-field, giving us a description of the in-trap dynamics (see \autoref{fig:fittedoscillations}~B) \cite{Manning:10}. 
Note that all frequency components of the potential in position space also arise in the velocity dynamics of that potential, with higher frequency components having a greater root-mean-square (rms) velocity. This is described further in \autoref{sm:damp_sine_vel}.

In metrology it is common to relate the performance of a measurement back to the expected uncertainty due to fundamental limitations (e.g. shot noise), however to date this has been rare in trap frequency measurements.
In \autoref{sm:stat_err} we derive the fundamental limiting performance of this method, assuming the measurement of the PAL position is limited purely by shot noise.

\subsection{Sampling rate limitation}
As our method relies on tracking the center-of-mass motion of the cloud, each atom laser pulse must be sufficiently separated in time in order to correctly group atoms to the pulse from which they originate, which sets a limit on the maximum rate that the BEC's motion can be sampled.
This maximum sampling rate decreases monotonically with atom number and trap frequency, due to the increased momentum width of the BEC. 
A derivation of the maximum sampling rate is provided in \autoref{sm:max_outc_freq}. 
For the parameters considered in this work, the maximum sampling rate is approximately $155$~Hz, well below many of the trap frequencies used in our experiment.
Thus, determining the true oscillation frequency practically requires a method that can use a sampling rate below that of the oscillation rate. To this end we show that there is a simple and elegant way around this problem by utilizing aliasing.

\subsection{Aliased signals}
\label{sec:aliased}

\begin{figure}[t]
\centering\includegraphics[width=0.8\textwidth]{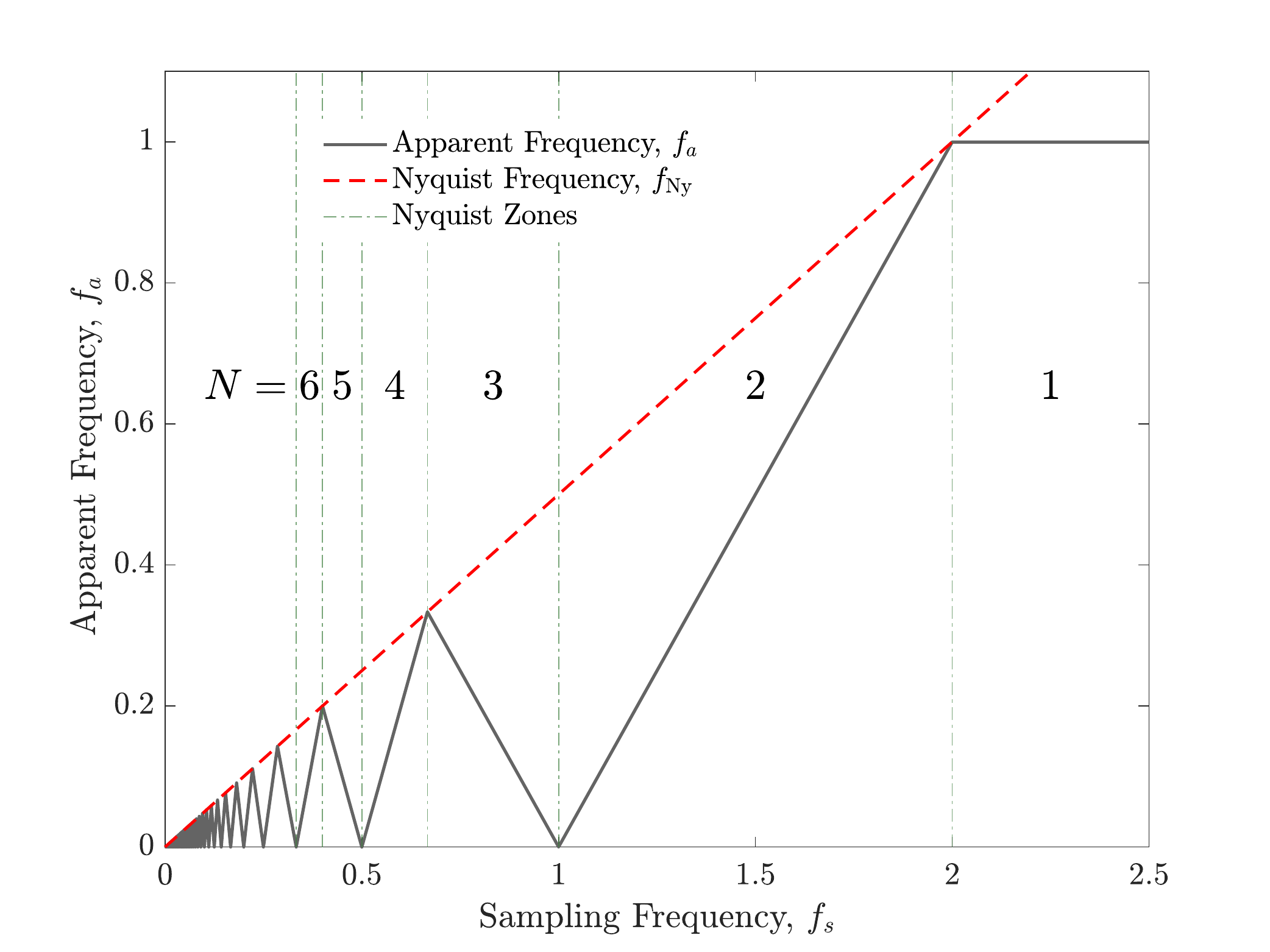}
\caption{
When a signal with a single frequency component ($f$) is aliased ($f_s<2f$) the apparent frequency (black solid line) changes as a function of the sampling frequency with an enveloped triangle wave shape.
This envelope is set by the Nyquist frequency (dashed red line) $f_{N}=f_{s}/2$ which is the maximum frequency that can be unambiguously determined at the sampling rate $f_s$ .
Green vertical dashed lines indicate edges of the first six Nyquist zones with the zone index indicated between them. Both axes are normalized by the true frequency $f$.
}
\label{fig:aliasedsignal}
\end{figure}

Given a continuous signal sampled instantaneously at a uniform rate $f_{s}$, then an interpolation of the sampled points will reconstruct the original continuous signal if it has no frequency components greater than half the sampling rate. This defines the the Nyquist frequency defined as $f_{N}=f_{s}/2$. 
However given arbitrary frequency content of the signal, then there are a family of possible signals that could produce the same sample series. 
When sampling a frequency component greater than $f_{N}$ that component will be \textit{aliased} and have an apparent frequency different to the true frequency, measured somewhere between zero and the Nyquist frequency.
To understand this process we use the concept of Nyquist zones, which are a division of the frequency spectrum into regions $f_{N}$ wide. 

The Nyquist zone of a signal of true frequency $f$ is given by:
\begin{equation}
N = \ceil{f/f_N},
\label{nzone}
\end{equation}
where $\ceil{x}$ is the ceiling function.
Under aliasing, the sampled data produces an apparent frequency $f_a$ inside the first Nyquist zone ($N=1$), which ranges in frequency from 0 to $f_{N}$.
The mapping can be seen as a folding of the input spectrum, as frequencies at even multiples of $f_{N}$ are mapped to zero while frequencies at odd multiples are mapped to $f_{N}$. In general the apparent frequency component $f_a$ is given by
\begin{equation}
  \label{reconstructingaliasing}
  f_a = \begin{cases} 
      -f +N f_s/2, & N \text{ even} \\
      f-(N-1)f_s/2, & N \text{ odd} \\ 
   \end{cases}
   \, ,
 \end{equation}
which is illustrated in \autoref{fig:aliasedsignal}. To find the true frequency unambiguously we need a method to find the Nyquist zone that the original signal lies in. In the relationship above the gradient of the apparent frequency with the sampling frequency $\partial f_{a}/\partial f_{s}$ uniquely specifies the Nyquist zone. 
Notably, this gradient only takes on integer values and therefore it may be measured with a relatively large uncertainty, so long as an integer value can be unambiguously determined.

In our application the reconstruction is simplified by two properties; firstly the motion of the BEC has a narrow frequency distribution which does not span multiple Nyquist zones. Secondly, given reasonable choice of sampling frequency,  the trap frequency is sufficiently stable that the oscillation will remain in the same Nyquist zone indefinitely.

\begin{figure}
\centering\includegraphics[width=0.8\textwidth]{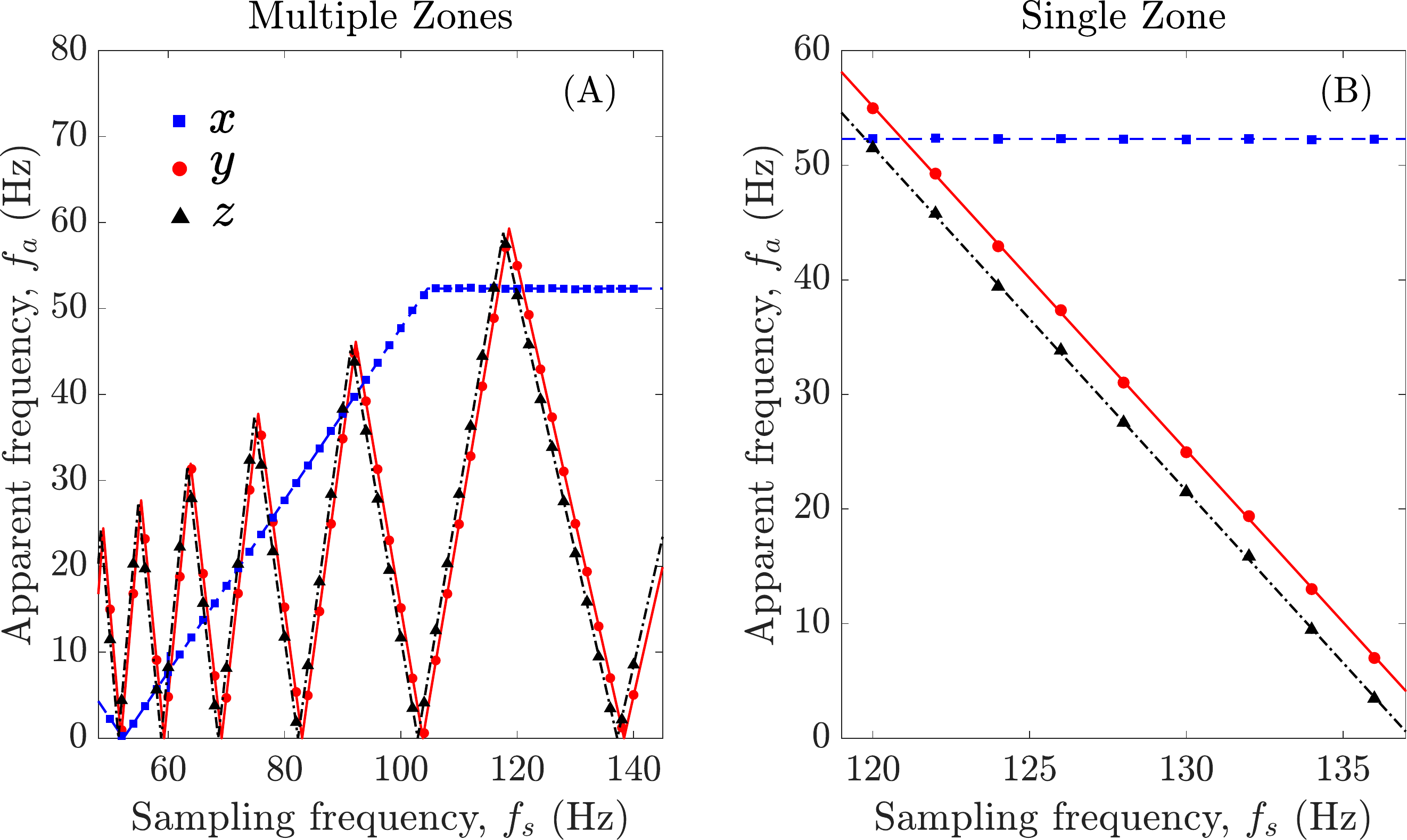}
\caption{Aliased frequencies across a range of sampling rates, with (A) displaying multiple Nyquist zones, and (B) showing the behavior within a single Nyquist zone. The fit in each dimension is indicated as follows: $x$, blue solid dashed line and squares; $y$, red solid line and circles; $z$, black dash-dotted line and triangles. Each point is the average of 3 measurements, with error bars showing standard deviation too small to see. Fits are of the form of \autoref{reconstructingaliasing}, with fit parameters for each dimension (A) $f_x=51.297(7)$~Hz, $f_y=411.67(1)$~Hz, $f_z=415.13(1)$~Hz (note these fits are across all zones), and (Right) $f_x=51.297(8)$~Hz, $f_y=411.67(2)$~Hz, $f_z=415.14(2)$~Hz.
}
\label{fig:aliasing}
\end{figure}

In our measurement of trap frequency we first conduct our atom laser method with a few sampling rates around a given frequency in order to characterize the Nyquist zone. 
This can be done by fitting a linear trend between $f_a$ and $f_s$ in a single Nyquist zone, with the gradient and intercept as fitting parameters that determine $N$ and $f$, respectively (see \autoref{fig:aliasing} (B)). 
Alternatively, the Nyquist saw-tooth function from \autoref{reconstructingaliasing} can be fit to data across multiple Nyquist zones, with the true frequency calculated as the lone fit parameter, as it determines the spacing of the Nyquist zones (see \autoref{fig:aliasing} (A) for example). 
In both examples in \autoref{fig:aliasing}, we see that when the Nyquist frequency becomes greater than the true frequency, as is the case for the $x$-axis frequency with $f_s>104$~Hz, the signal is no longer aliased, and we observe a flat relationship between apparent and sampling frequency. 
Importantly, as the uncertainty in $f_s$ is negligible ($<0.2$~ppm), once the Nyquist zone is determined the uncertainty of the true frequency is given solely by the uncertainty in the apparent frequency, allowing for very high relative precision. 
In this work we focus only on reconstructing signals with a single frequency component for simplicity, however it may be generalised to more complex signals by considering the aliased value of each frequency component individually \cite{ChunLinphd,2017arXiv170505216D}.

\section{Results}
The method as described above may be applied in all three dimensions simultaneously however here we will concentrate on the $y$ direction (see \autoref{fig:fittedoscillations} (A)), in a tight axis of our cylindrically symmetric trap. From a single outcoupling sequence, the detector provides the average position of the condensate as a function of time in the far field, allowing us to reconstruct the in-trap velocity. We can then determine the apparent frequency of oscillation by fitting a damped sine wave to the in-trap velocity:
\begin{equation}
\label{model}
v(t) = A e^{-\lambda t} 2 \pi f_{d} \sin(2\pi f_a t+\phi ) + c ,
\end{equation}
where $\lambda$ is the damping rate, $A$ is the amplitude, $f_a$ is the apparent frequency (in Hz, restricted to the range $[0,f_s/2]$), $\phi$ is the phase of the oscillation at $t=0$, and $c$ is an offset to compensate for small offsets in the velocity reconstruction.
Alternatively, we can use the discrete Fourier transform of the raw data to determine the apparent frequency (\autoref{fig:fittedoscillations} Right), and use the signal-to-noise ratio of the peak to determine the upper limit of the uncertainty (see\autoref{sec:fourier_unc} and \cite{refld0}). This gives us an independent measure of value and uncertainty of the apparent frequency, and importantly serves as a strong validation and benchmark of the underlying single-shot uncertainty.

A comparison of the fitted model to the raw data and Fourier transform of a single-shot is shown in \autoref{fig:fittedoscillations}. The fitted function returns an estimate of the standard error for each parameter, calculated through the summation of the squared residuals. The fitted frequency of the sine wave, $f_a = 2.998(6)$~Hz, has an associated error of $6$~mHz, compared to the peak Fourier component of $=2.996(13)$~Hz, where the uncertainty is inversely proportional to the interrogation time and signal-to-noise ratio \cite{refld0}. As both values are within error of one another, and have similar associated uncertainties, we can be confident in the magnitude of the single-shot uncertainty.

To determine the Nyquist zone we measure the apparent frequency across a range of sampling frequencies. \autoref{fig:aliasing} shows typical results with our Nyquist saw-tooth fit (\autoref{reconstructingaliasing}) across both a wide and narrow range of sampling frequencies. The fit parameter corresponding to true frequency here is given by $f_x=51.297(7)$~Hz, $f_y=411.67(1)$~Hz, $f_z=415.13(1)$~Hz.
If the variation in the true frequency is less than the width of a given Nyquist zone, we can then invert \autoref{reconstructingaliasing} to obtain the true frequency. As the stability of our trap is on the order of millihertz, we only need to perform this aliasing procedure once for each trap configuration. Once we determine the Nyquist zone we can use a single sampling frequency to measure changes in frequency, for example from frequency drift, or due to applied potentials for sensing purposes.

\subsection{Trap stability}

\begin{figure}
\centering
\includegraphics[width=\textwidth]{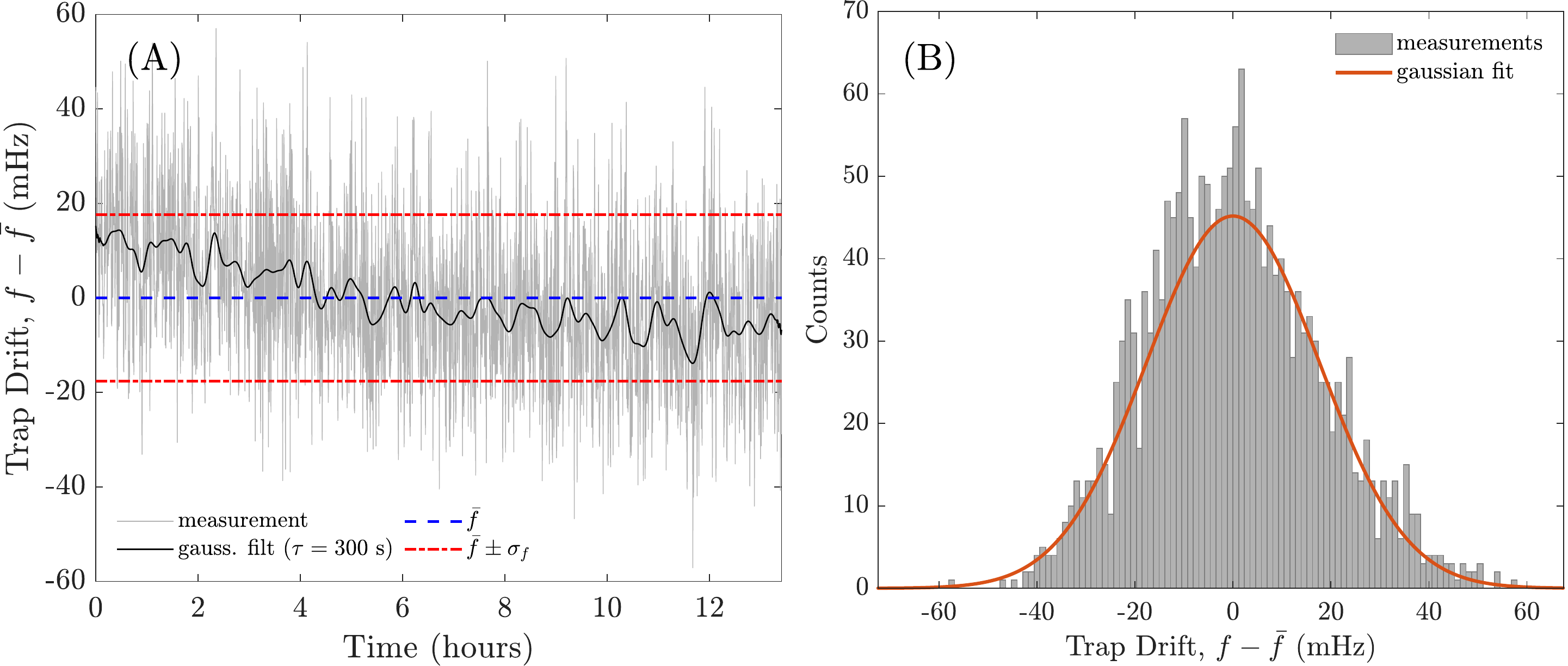}
\caption{Drift of measured trap frequency for 1929 experimental runs taken over $\sim 13$ hours. (A) Time series of difference between measured trap frequency and the average, with dot dashed red and dashed blue lines indicating standard deviation and mean of the series respectively. Single-shot uncertainties are not shown for clarity. The time between measurements is set by the 24~s shot duration.
(B) Histogram of measured values, solid orange line is a Gaussian fit to the data with standard deviation of $17.6$~mHz.
}
\label{fig:drift}
\end{figure}

As a benchmark of our method we quantify the stability of our magnetic trap.  In our experiment the trap currents are stabilized with a current regulator to a fractional stability of $\approx50$~ppm and background fields are actively stabilized to $\approx3$~nT \cite{Dedman2007}.
The trap frequency measured in the $y$-axis over $\approx2000$ measurements is shown in \autoref{fig:drift}. Over $\approx$13 hours we observe a drift of order of tens of millihertz ($50$~ppm) and an approximately Gaussian distribution of measurements.

\begin{figure}
\centering
\includegraphics[width=0.7\textwidth]{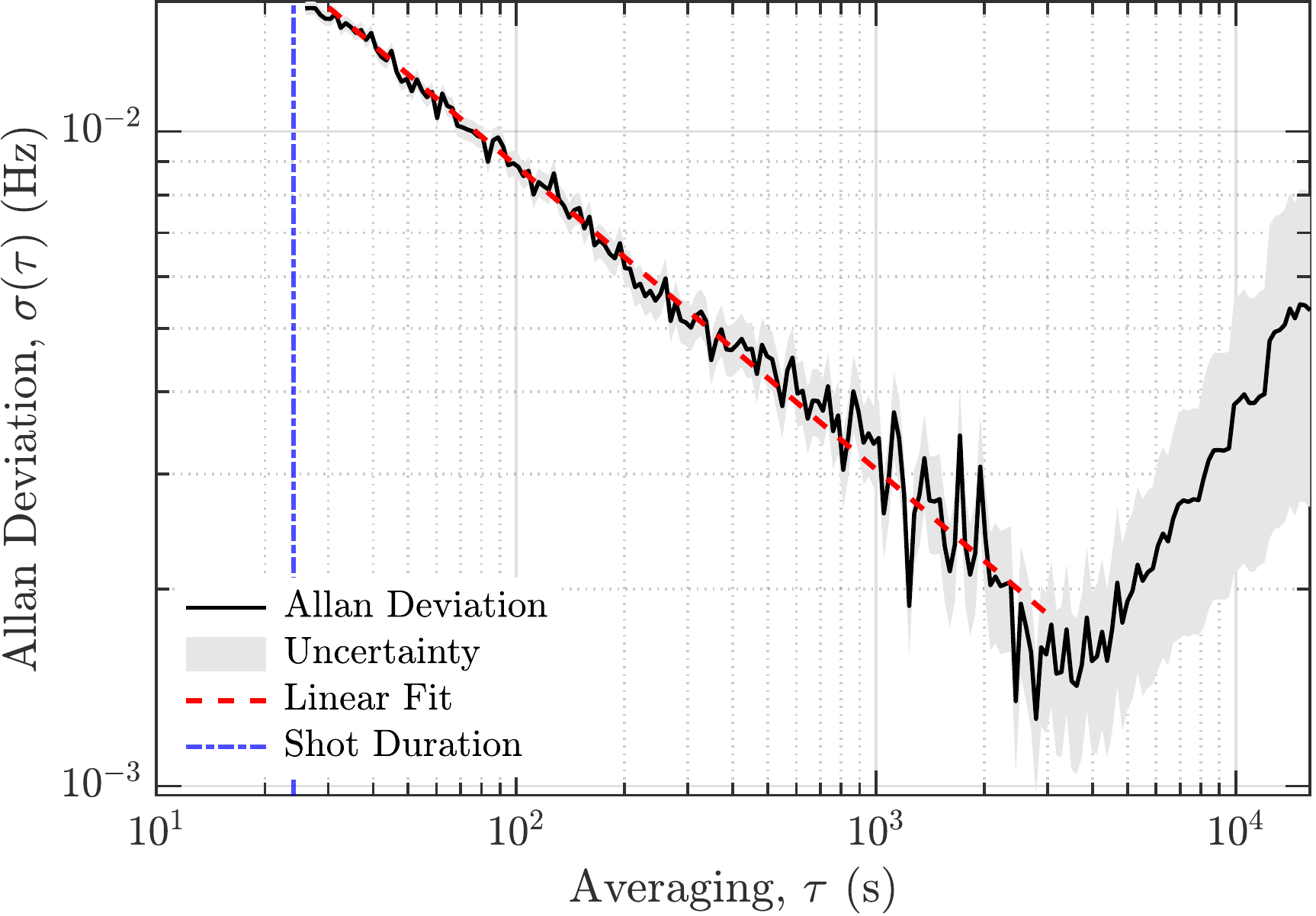}
\caption{Allan deviation of the $y$-axis frequency of our experimental magnetic trap, using the same data set as presented in \autoref{fig:drift}. 
The shot duration (24~s) which sets the time between measurements is indicated with the vertical dot-dashed line. 
A fit of the form $1/\sqrt{\tau}$, for $\tau$ in the range $[0,3000]$ seconds, is shown as a red dashed line, with a proportionality constant of $15.7$~mHz/$\sqrt{\text{Shots}}$. 
The deviation reaches a minimum of $1.5$~mHz at around 2 hours. At time scales longer than 2 hours the deviation begins to increase due to long term systematic drifts, such as temperature changes of the trap current sense resistor.
}
\label{fig:allan}
\end{figure}

To give a quantitative measure of our measurement noise we find the Allan deviation $\sigma(\tau)$ of successive measurements (\autoref{fig:allan}). 
The minimum of the Allan deviation is $\sigma(\tau)=1.4$~mHz (3.4~ppm) and occurs at an averaging time of $\tau=2700$~s $=0.75$~hrs. At times less than this we find that the Allan deviation is given by $\sigma=86\mathrm{~mHz}/\sqrt{\mathrm{Hz}}$ which for our shot duration of $\approx 24$~s corresponds to $\sigma=16\mathrm{~mHz}/\sqrt{\mathrm{Shots}}$. 
This deviation corresponds to a trap frequency measurement with a fractional uncertainty of 39~ppm in a single-shot, improving on the current record in the literature by a factor of 3 \cite{Harberphd}.

\subsection{Statistical and systematic error}

\begin{figure}
\centering
\includegraphics[width=0.5\textwidth]{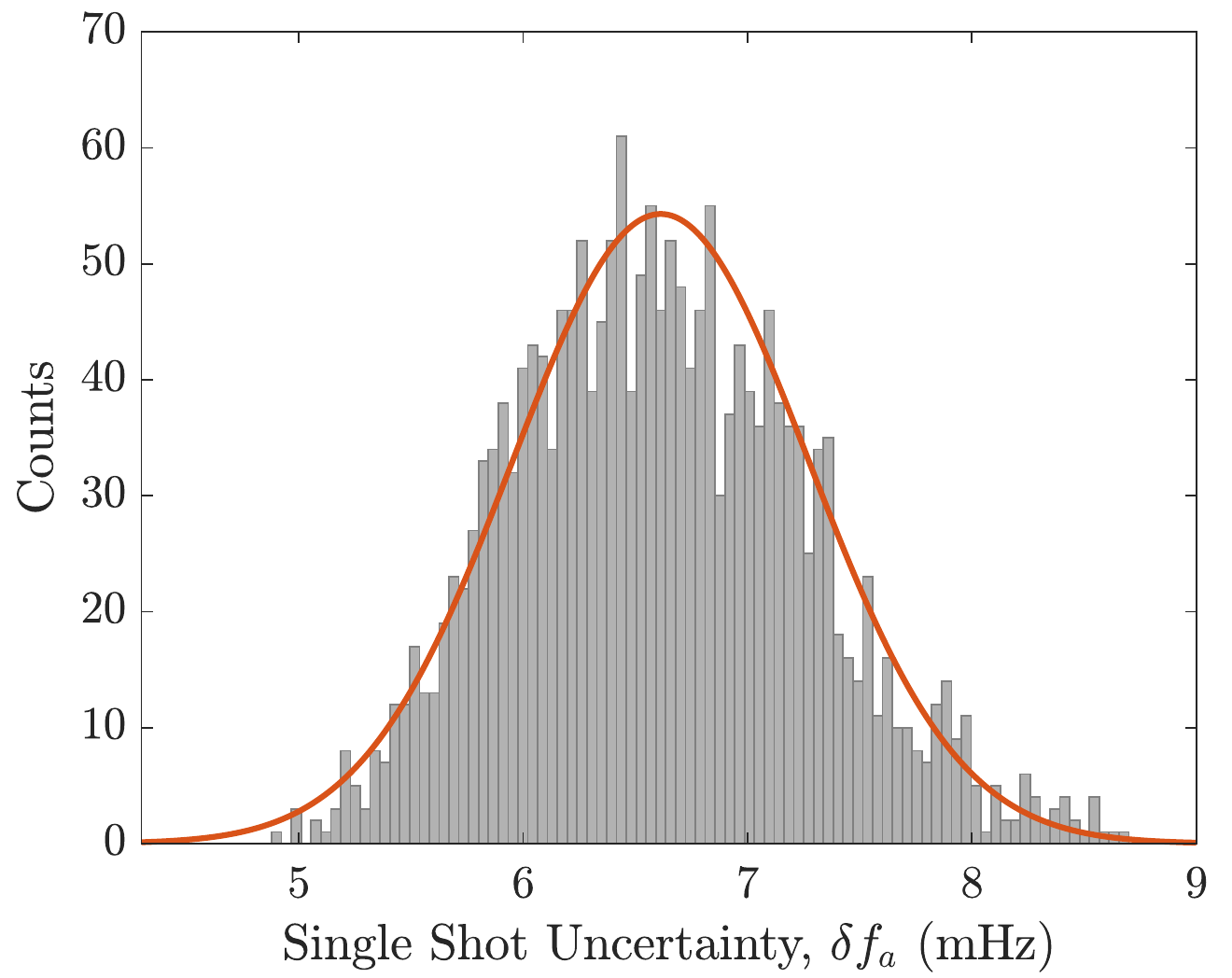}
\caption{Histogram of single-shot uncertainty, approximately 2000 experimental runs. We perform our sampling in the third Nyquist zone. The Gaussian fit, solid orange line, has a mean of $6.6$~mHz, and a standard deviation of $0.66$~mHz.
}
\label{sm:fig:single_shot_unc}
\end{figure}

The two factors which dominate the uncertainty of this method are the damping rate of the oscillation and the standard error of the PAL mean velocity. The former limits the interrogation duration ($T$) which would otherwise scale the frequency uncertainty ($\sigma_\omega$) as $1/T$ and is increased by the thermal fraction, trap frequency and anharmonic components of the trap. The standard error of the PAL mean velocity is determined by the velocity distribution of an atom laser pulse in the far field which arises from the complex outcoupling dynamics we have investigated previously \cite{PhysRevA.97.063601}. 
We have derived a simple expression for this uncertainty given in \autoref{sm:stat_err} which has been validated with simulations of outcoupling dynamics, both predicting a single-shot uncertainty of $\approx5$~mHz. 
Considering the discrepancy between this value and the measured single-shot deviation we consider it likely that the variations in the trap current dominate the measured deviation.
Given the measured fractional trap current noise (50~ppm) and the square root relationship between current and trap frequency, then the expected fractional frequency stability is half the fractional trap current noise (25~ppm), in approximate agreement with the measured deviation.
Considering the estimated standard error of the fit frequency from the individual fits (6.6(7)~mHz \autoref{sm:fig:single_shot_unc}) we expect that this current noise is small on the timescale of a single measurement (1.2~s) and dominates shot to shot deviation.
With improved current regulator stability \cite{colincurrentctr} we expect that an uncertainty near the predicted value of 5~mHz (12~ppm) would be possible with this trap configuration. If further improvements are sought then it is possible to use a trap with much lower frequencies which decreases both the damping rate of oscillations and the velocity width of the PAL.

We have also quantified the small systematic shift in the measured oscillation frequency ($\approx$-80~mHz) caused by the anharmonic shape of the magnetic trap \autoref{sec:anh_shift}.

\section{Discussion}

We have developed a method for measuring trap frequency in 3-dimensions simultaneously with an uncertainty better than the state of the art.
This method has formed a key tool in many of our recent works;
In \cite{henson2021precision} it was employed to find the trap frequency of a probe beam overlapped with the magnetic trap. This formed a sensitive measurement of the optical dipole polarizability of helium and in turn the \textit{tune-out} frequency, at which the dipole polarizability vanishes, as a stringent test of QED.
The ability to measure the trap frequency of multiple axes simultaneously allowed for quick calibration in \cite{rossQD} where it was used to quantitatively compare theoretical predictions with experimental measurements.
In \cite{Henson13216} this method was used to find the amplitude of oscillations induced by transformation of a BEC trapping potential, in order to optimize the transformation process.

In this work we have considered the uncertainty for a single shot (realization of a BEC) in order to compare with other works with different BEC preparation times. 
However, as most experiments average the results of many single shot measurements the total uncertainty is also inversely proportional to the square root of the number of measurements per unit time, which is generally dominated by BEC preparation time. 
By combining the method described here with the approaches for rapid BEC creation (eg. 3.3~s \cite{PhysRevA.103.053317} for He*) then a factor of \(\sim 2\) improvement could be made in the measurement uncertainty of an experiment with fixed duration.

While we have focused on trap frequency the method also produces an accurate measurement of the oscillation amplitude and phase, both of which have interesting applications. 
For example, calibrating the quantity of energy transferred to a system via the center of mass motion for calorimetry experiments. The non-destructive nature of this method also opens the possibility to use the same BEC for both trap frequency measurement and further experimentation. We have used this approach, with reduced outcoupling fraction, in some experiments to save time and improve uncertainty.

At present the strong interaction between outcoupled atoms and the BEC cause the PAL to have a large velocity width. It may however, be possible to reduce this width and improve the fundamental uncertainty using a Raman outcoupling process which out-couples atoms with an initial (downwards) velocity \cite{Debs2009}.
This initial velocity reduces deflection of the outcoupled atoms by the BEC and thus reduces the velocity width in the far field. 
A further advantage of Raman outcoupling would be that it could be used to measure the trap frequencies of non-magnetic traps, such as optical dipole traps, where all $m_{J}$ sub-states are trapped and thus RF outcoupling is not possible.

In this work we make heavy use of single atom detection of He*'s for repeated measurement of the PAL in the far field.
In principle, the method outlined in this work may be adapted to other atomic species (for example Rubidium, Cesium, Potassium, \textit{ect.}) using a \textit{light-sheet} imaging system \cite{lightsheetSchmidemayer}. 
While using a BEC is preferable due to its narrow momentum width and low oscillation damping rate, this method could also be used to measure the trap frequency of trapped thermal Bosons or cold Fermions.

To summarize, we have outlined a novel method to measure trap frequency of a confined BEC. We use a PAL to sample the velocity of an oscillating BEC, allowing us to decrease the fractional uncertainty to 39~ppm in a single experimental realization. Additionally, we demonstrate a reconstructive aliasing approach to allow trap frequencies much higher than the sampling frequency to be measured. 

\section*{Funding}
This  work  was  supported  through Australian  Research  Council  (ARC)  Discovery  Project Grants No.  DP160102337, DP180101093 and DP190103021.
K.F.T. was  supported  by  Australian  Government  Research Training  Program  (RTP)  scholarships.
S.S.H.  was supported  by  ARC  Discovery  Early  Career  Researcher Award  No.  DE150100315.

\section{Appendix}

\subsection{Atom Laser Power Ramp }
If the pulsed atom laser is driven with a constant RF power then the atom number in each pulse will proceed as a geometric series which increases the position uncertainty of later pulses and gives a larger overall uncertainty.
To partially correct for this effect we increase the RF power with increasing pulse number which gives a more uniform number per pulse.


\subsection{Damped Sine Wave In Velocity}
\label{sm:damp_sine_vel}
To find the trap frequency from repeated measurements of the BEC velocity we require a model of the dynamics.
The low excitation energies present in this experiment mean that the harmonic approximation is a good model of the potential about its minimum.
The motion of an oscillating BEC is damped by the redistribution of energy from the BEC oscillation into the thermal cloud (Landau damping) through scattering along with more complex hydrodynamic effects \cite{PhysRevLett.80.2269,Yuen_2015}. 
While predicting the damping rate is possible, for our use it is sufficient to merely observe that the damping rate will increase with increasing trap frequency and temperature of the thermal component.
Therefore a reasonable model for the center of mass position $x(t)$ of the BEC during oscillations is the classical damped harmonic oscillator. The time dependence is then given by
\begin{equation}
    \label{sm:eq:damp_sine_vel}
    x(t)=A e^{-t\lambda} \sin{(2 \pi f_{d} t+\phi)}
\end{equation}
where $A$ is the initial oscillation amplitude, $\lambda$ is the damping rate. Here $f_{d}$ is the dampened frequency given as
\begin{equation}
   f_{d}=f_{0}\sqrt{1-\zeta^2}
\end{equation}
where $f_{0}$ is the undamped frequency, and $\zeta$ is the damping ratio which is given by 
\begin{equation}
   \zeta = \frac{\lambda}{\sqrt{\lambda^2+(2\pi f_{0})^2}} .
\end{equation}
For the conditions used in this work the $ f_{d}$ may be approximated as $ f_{0}$ within a fractional error of $\approx10^{-8}$.
The pulsed atom laser used in this work measures the velocity component of this oscillation, therefore we find the derivative of \autoref{sm:eq:damp_sine_vel} to be
\begin{align}
    \derivative{x(t)}{t}& = v_x(t) =A e^{-\lambda t}\left( 2\pi f_{d} \cos{(2\pi f_{d} t+\phi)} - \lambda \sin{(2\pi f_{d} t+\phi)}\right).
\end{align}
This can be simplified by using the sine cosine summation identity \cite{HarmonicAdd} to give an expression as a single sine wave with a phase and amplitude shift 
\begin{align}
    v_x(t) = & A e^{-\lambda t}
        \sqrt{(2\pi f_{d})^2 + \lambda^2} 
        \sin{\left(2\pi f_{d} t+\phi-
            \arctan{\left( \frac{-\lambda}{2\pi f_{d}} \right)}
        \right)} \\
        \approx & A e^{-\lambda t}
        2 \pi f_{d}
        \sin{( 2 \pi f_{d} t+\phi)}.
        \label{eq:com_bec_motion}
\end{align}
For the conditions used in this work the phase shift ($\arctan{( -\lambda / \omega_{d} )}$) is negligible ($\approx5\times10^{-4}$ rad) and the amplitude term can be approximated $\sqrt{\omega_{d}^2 + \lambda^2}\approx\omega_{d}$ within a fractional error of $\approx10^{-7}$. Therefore we have a simple model of the dynamics which we can use as a fit function of the PAL velocity in order to extract the trap frequency of the confining potential.

\subsection{Outcoupling Dynamics}
\label{sm:outc_sim}
We have previously shown that for strongly confining cigar-shaped traps used here that the final velocity distribution of outcoupled atoms is governed by an initial repulsion from the BEC in the tight axes followed by a collision with the BEC as they fall \cite{PhysRevA.97.063601}. This generates a disk shaped distribution in the tight axes (here $y$, $z$) along with a vertical cut upwards and a number of interference fringes. 

While the dynamics of the pulsed atom laser require a full quantum treatment to predict the entire structure a semi-classical model can accurately predict many features of the velocity distribution. This treatment neglects interactions between atoms in the atom laser, which we have found to be negligible previously. As it is a classical simulation it is unable to reproduce the interference fringes in the density profile, however they have negligible impact on the density profile when integrated to one dimension.

We start with a Thomas-Fermi approximation for the BEC's density and assume its center of mass motion can be approximated as a dampened sine wave \autoref{eq:com_bec_motion}.
To simulate the outcoupled atoms we sample starting positions from the BEC's density distribution and set the velocity at rest with respect to the mean BEC velocity. 
We then integrate classical equations of motion under gravity and the potential generated by the mean field from the moving BEC. 
We continue this integration until the atoms have expanded sufficiently that they do not interact with the BEC. 

Using the Thomas Fermi approximation for a BEC from \cite{Pethick} we find the mean field potential acting on an atom in the atom laser from the original BEC, ($ V_{mf} $) at position $ \mathbf{r} $ to be
\begin{equation}
V_{mf}(\mathbf{r})=
\begin{cases} 
      \frac{U_{ac}}{U_{cc}}\left( \mu-V(\mathbf{r}) \right) & V(\mathbf{r})<\mu \\
      0 & V(\mathbf{r})\geq \mu \\
   \end{cases}
\end{equation}
where the chemical potential $\mu$ is defined as
\begin{equation}
    \mu=\frac{15^{2/5}}{2} \left(\frac{N a}{\bar{a}}\right)^{2/5} \hbar \bar{\omega}
\end{equation}
and
\begin{equation}
    \bar{a}=\sqrt{\frac{\hbar}{m \bar{\omega}}}
\end{equation}
\begin{equation}
    \bar{\omega}=\left( \omega_{x}  \omega_{y} \omega_{z} \right)^{1/3}.
\end{equation}
The terms $ U_{ac}$,$ U_{cc}$ are the interaction strengths between the condensate, and between atoms in the condensate and atoms in the atom laser respectively. For helium these interaction strengths are equal $ U_{ac} = U_{cc} = 4 \pi \hbar^2 a /m$ with scattering length $a=7.51$~nm \cite{PhysRevLett.96.023203}.

The maximum velocity that a classical atom can reach is by converting the maximum mean field energy into kinetic energy, corresponding to an atom starting at the center of the BEC. This corresponds to a velocity of
\begin{equation}
    v_{max}=\sqrt{\frac{2 \mu}{m}}
\end{equation}
which defines the radius of the PAL disk.

\subsection{Maximum Outcoupling Frequency}
\label{sm:max_outc_freq}
We wish to calculate the maximum outcoupling frequency at which two subsequent atom laser pulses do not overlap one another at detection. Consider the upward and downward going atoms generated by outcoupling $v_{z}=\pm v_{max}$, our criteria for overlap is when the down going trajectory $v=+v_{max}$ from pulse $n$ overlaps with the up-going trajectory $v=+v_{max}$ from pulse $n+1$. This case is simply the time for the up-going trajectory from pulse $n$ to come back down to the original BEC, 
$ t_s=2 v_{max}/g$, where g is the acceleration due to gravity.

This treatment neglects that some fraction of the up-going atoms from an atom laser pulse scatter off the BEC as they fall generating Bogoliubov-Cherenkov (BCR) radiation tails \cite{PhysRevA.97.063601}, this scattering is bounded by a reflection in the vertical direction (in principle, there is a secondary scattering process, however its probability is negligible). To be sure that there is no overlap of these scattered trajectories then an outcoupling period of 
$ t_s=4 v_{max}/g$
is necessary.  
We have found that an intermediate between these cases is sufficient as the BCR tails that extend upwards quickly drop in density compared to the bulk, in general we have used $ t_s\approx3 v_{max}/g$ as the fastest practical sampling.

\begin{equation}
    f_{s,max}= \frac{1}{2} \frac{g}{v_{max}} =  \frac{g}{3}  \sqrt{\frac{m}{2 \mu}}
\end{equation}

For our experimental conditions
$(\omega_x,\omega_y,\omega_z) \approx 2\pi(51,412,415) $~Hz, 
$N=5\times10^{5}$, and 
$g=9.8\mathrm{~m}/\mathrm{s}^2$ we have 
$f_{s,max}\approx155$~Hz which is in good agreement with experiment.

\subsection{Statistical error}
\label{sm:stat_err}
To provide a scale to compare experimentally measured performance against, we derive an analytical expression for the fundamental limit set by shot noise on the standard error in the trap frequency found using the PAL method. This investigation will also serve to inform researchers looking to use this method as part of a sensing methodology or system characterization.

The dominant error source that remains after technical noise sources (eg. trap current noise) are eliminated is the error in determining the mean velocity of an atom laser pulse. This shot noise in the mean velocity is determined by the finite number of detected atoms per pulse and the velocity distribution arising from the outcoupling dynamics. 
For an analytical treatment of the error we will first approximate the the one-dimensional density distribution of a PAL along the tight axes of a cigar shaped trap as a uniform distribution between $\pm v_{max}$. As will be shown later this approximation gives excellent agreement to more detailed simulations.

Under this approximation the probability distribution of the atom laser pulse mean position follows a Bates distribution with corresponding standard error of,
\begin{equation}
\label{eq:sigma_al}
\sigma_{v}=\sqrt{\frac{1}{12 n_{p}}} 2 v_{max} 
\end{equation}
where $n_{p}$ is the number of detected atoms for a single pulse. 

While a fit to a dampened sine wave with noisy samples is common across many disciplines an analytic equation predicting the standard error in the fit frequency cannot be found in the literature. To correct this gap we have extended the treatment for the standard error in fit parameters from an undampened sine wave given by \cite{Montgomery1999}, to the exponentially enveloped case and find the standard error in the fit trap frequency $\sigma_{f}$ as:     
\begin{equation}
\label{eq:theory_f_unc_damp_sine}
\sigma_{f}= \frac{\sigma_v}{A_v}  \frac{1}{\sqrt{M \pi}}  \frac{4}{5}
\sqrt{\frac{  T \lambda^{3} (-1 +e^{2 T \lambda} )  }{-1-2 T^2 \lambda^2 + \cosh{(2 T \lambda)} }}
,
\end{equation}
where $M$ is the number of samples of the dampened sine wave, $T$ is the total interrogation duration, $A_v$ is the amplitude of the oscillation, $\sigma_{v}$ is the noise in each sample and $\lambda$ is the damping rate. Within the commonly used parameter domains of: signal to noise $\sigma_v/A_v<1$, measurement duration relative to oscillation period $T>30/\omega$ and weakly damped oscillations $1/\lambda> 30 \omega$, this expression gives results that are within 20\% of repeated numerical experiments.
This lengthy derivation is outside the scope of this work and will be detailed in an upcoming paper.

In applying \autoref{eq:theory_f_unc_damp_sine} we assume that the standard error of each sample is the same. While in our experiment the standard error of each velocity measurement, set by the mean field repulsion from the original BEC, strictly depends on atom number this dependence is sufficiently weak that it may be
ignored\footnote{The ratio of widths for a pulse removed from a BEC which has an atom number 10\% of the starting value is a factor of 0.63 smaller. 
For a series of PALs where the atom number linearly decreases to a final atom number 10\% of the starting value, the average spatial width of the pulses will be 0.87 that of the first.}.
In the case of a fixed sample number (c.f a fixed sample rate) then the minimum standard error in frequency is achieved at $T\approx2/\lambda$ giving an uncertainty  of:
\begin{align}
\sigma_{f}&\approx 1.092\frac{ \sigma_x \lambda}{A_v \sqrt{M}} 
\approx 0.631 \frac{ v_{max}  \lambda}{A_v \sqrt{N \eta}} 
\end{align}
where $\eta$ is the quantum efficiency of the detector.
For the work presented here the measurement duration was shorter than this optimum condition and  \autoref{eq:theory_f_unc_damp_sine} is used. For the conditions of our experiment $\lambda=0.30$, $n_{p}\approx257$, $M=156$, $A_v=10$~mm/s, and $v_{max}\approx 24$~mm/s this gives a standard error in the trap frequency of $\approx 5$~mHz. 
To validate this approach we use a semi-classical simulation of the outcoupling dynamics (as above) combined with a damped sine wave fit to the mean pulse position to give the estimated trap frequency. By repeating this simulation many times we reach a estimate of the standard error in the trap frequency due to the finite atom number in each AL pulse of $\sigma_f = 5.0(4) $~mHz.
This shows that the basic approximations made for the noise in the system produce a reasonable estimate for the uncertainty and validate this expression for future use.

\subsection{Spectral Uncertainty}
\label{sec:fourier_unc}

Following the procedure of \cite{refld0} we can derive an uncertainty in the underlying frequency from the DFT amplitude spectrum. Importantly, this is a basic test and is completely separate from the uncertainty derived from the least squared sinusoidal fit (primarily stemming from this method assuming no underlying functional form of the signal). In this work they heuristically derive an upper limit on the frequency error based on a model of mono-periodic signals with white noise and validate this expression with numerical tests.

The uncertainty criterion is inversely proportional to the interrogation time $T$, and the square root of spectral significance $\sqrt{sig}$. The spectral significance is defined as one on the logarithm of the probability that a DFT peak arises purely due to noise. We can also express the relation in terms of the signal-to-noise ratio $S/R$ of the DFT peak, which is defined as the peak amplitude divided by the average amplitude in a given frequency range.
\begin{align}
    \sigma_{f,DFT} &= \frac{1}{T \sqrt{sig}} = \frac{4}{\pi \log (e) T S/R}
\end{align}
We further validate both this technique and the least square fit uncertainty using numerical experiments.

\subsection{Anharmonic Frequency Shift}
\label{sec:anh_shift}
\begin{figure}
\centering
\includegraphics[width=0.7\textwidth]{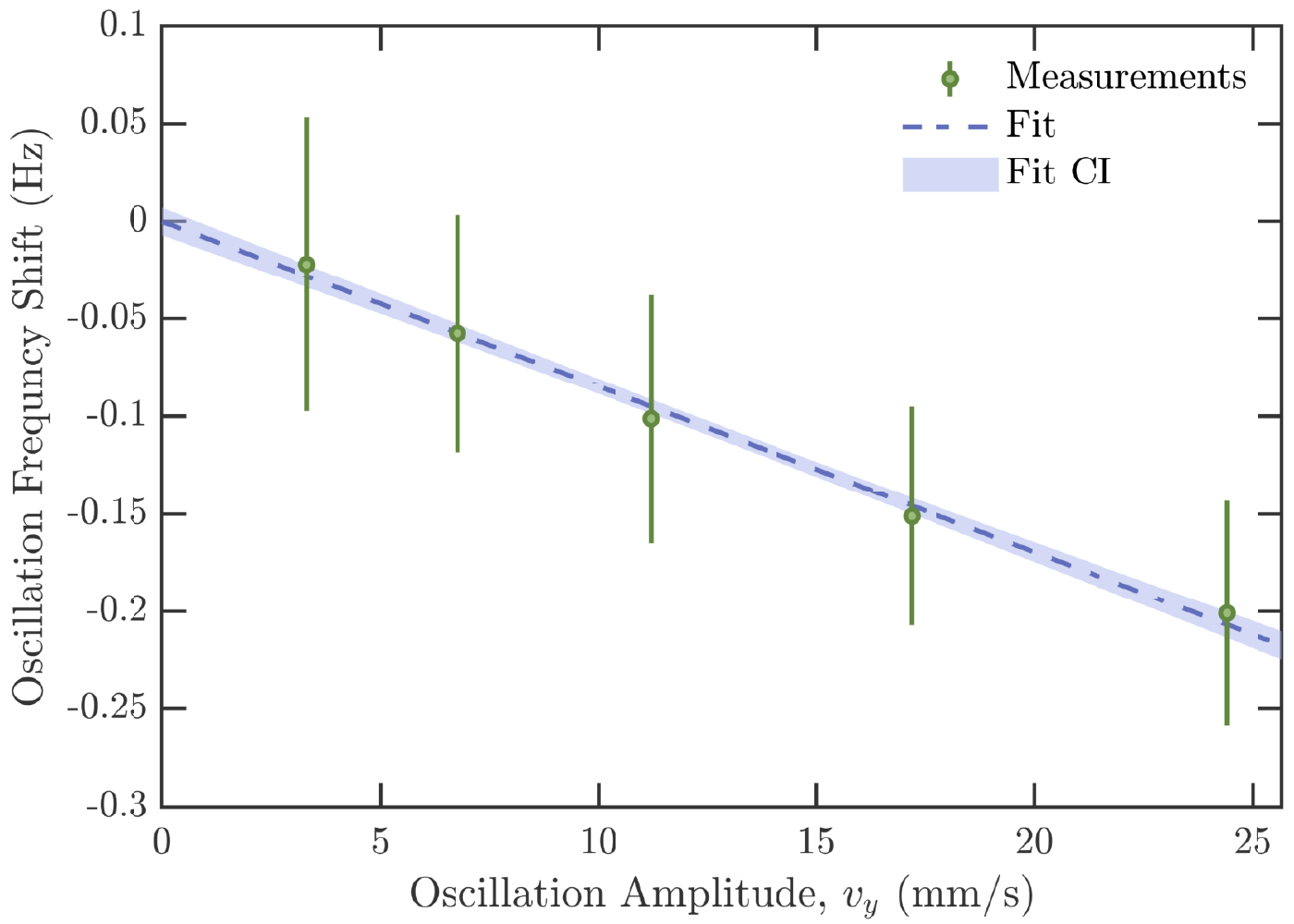}
\caption{Oscillation frequency (green round markers) measured as a function of oscillation amplitude, showing the weak anharmonicity of this trap.
A linear fit to these measurements (dashed blue line) has a gradient of $-8.4(4) \: \mathrm{Hz}\cdot\mathrm{m}^{-1}\cdot\mathrm{s}$.
The error bars show the standard error in the measured trap frequency averaged over hundreds of measurements, while the shaded region shows the confidence interval in the fit. 
}
\label{sm:fig:anh_shift}
\end{figure}

While a reasonable approximation, the potential of the magnetic trap used in this work is not strictly harmonic.
The primary effect of this non ideal \textit{anharmonic} shape is to change the trap frequency as a function of oscillation amplitude. 
To study this effect we have repeated trap frequency measurements at a range of oscillation amplitudes by varying the duration of the magnetic field gradient used to set the BEC in motion.
The resulting oscillation frequency shows a approximately linear decrease with oscillation amplitude, see \autoref{sm:fig:anh_shift}. 
This dependence shows that the observed trap frequency measured with an oscillation amplitude of 10~mm/s (as used in the main text) is decreased by 80(10)~mHz compared to the zero amplitude limit.

\bibliography{TrapFreqMet.bib}

\begin{thebibliography}{51}%
\makeatletter
\providecommand \@ifxundefined [1]{%
 \@ifx{#1\undefined}
}%
\providecommand \@ifnum [1]{%
 \ifnum #1\expandafter \@firstoftwo
 \else \expandafter \@secondoftwo
 \fi
}%
\providecommand \@ifx [1]{%
 \ifx #1\expandafter \@firstoftwo
 \else \expandafter \@secondoftwo
 \fi
}%
\providecommand \natexlab [1]{#1}%
\providecommand \enquote  [1]{``#1''}%
\providecommand \bibnamefont  [1]{#1}%
\providecommand \bibfnamefont [1]{#1}%
\providecommand \citenamefont [1]{#1}%
\providecommand \href@noop [0]{\@secondoftwo}%
\providecommand \href [0]{\begingroup \@sanitize@url \@href}%
\providecommand \@href[1]{\@@startlink{#1}\@@href}%
\providecommand \@@href[1]{\endgroup#1\@@endlink}%
\providecommand \@sanitize@url [0]{\catcode `\\12\catcode `\$12\catcode
  `\&12\catcode `\#12\catcode `\^12\catcode `\_12\catcode `\%12\relax}%
\providecommand \@@startlink[1]{}%
\providecommand \@@endlink[0]{}%
\providecommand \url  [0]{\begingroup\@sanitize@url \@url }%
\providecommand \@url [1]{\endgroup\@href {#1}{\urlprefix }}%
\providecommand \urlprefix  [0]{URL }%
\providecommand \Eprint [0]{\href }%
\providecommand \doibase [0]{https://doi.org/}%
\providecommand \selectlanguage [0]{\@gobble}%
\providecommand \bibinfo  [0]{\@secondoftwo}%
\providecommand \bibfield  [0]{\@secondoftwo}%
\providecommand \translation [1]{[#1]}%
\providecommand \BibitemOpen [0]{}%
\providecommand \bibitemStop [0]{}%
\providecommand \bibitemNoStop [0]{.\EOS\space}%
\providecommand \EOS [0]{\spacefactor3000\relax}%
\providecommand \BibitemShut  [1]{\csname bibitem#1\endcsname}%
\let\auto@bib@innerbib\@empty
\bibitem [{\citenamefont {Pethick}\ and\ \citenamefont
  {Smith}(2008)}]{Pethick2008}%
  \BibitemOpen
  \bibfield  {author} {\bibinfo {author} {\bibfnamefont {C.~J.}\ \bibnamefont
  {Pethick}}\ and\ \bibinfo {author} {\bibfnamefont {H.}~\bibnamefont
  {Smith}},\ }\href {https://doi.org/10.1017/cbo9780511802850} {\emph {\bibinfo
  {title} {Bose{\textendash}Einstein Condensation in Dilute Gases}}}\ (\bibinfo
   {publisher} {Cambridge University Press},\ \bibinfo {year}
  {2008})\BibitemShut {NoStop}%
\bibitem [{\citenamefont {Harber}(2005)}]{Harberphd}%
  \BibitemOpen
  \bibfield  {author} {\bibinfo {author} {\bibfnamefont {D.~M.}\ \bibnamefont
  {Harber}},\ }\emph {\bibinfo {title} {Experimental investigation of
  interactions between ultracold atoms and room-temperature surfaces}},\ \href
  {https://jila.colorado.edu/sites/default/files/2019-05/harber_thesis.pdf}
  {Ph.D. thesis},\ \bibinfo  {school} {University of Colorado} (\bibinfo {year}
  {2005})\BibitemShut {NoStop}%
\bibitem [{\citenamefont {Bertoldi}\ \emph {et~al.}(2019)\citenamefont
  {Bertoldi}, \citenamefont {Minardi},\ and\ \citenamefont
  {Prevedelli}}]{PhysRevA.99.033619}%
  \BibitemOpen
  \bibfield  {author} {\bibinfo {author} {\bibfnamefont {A.}~\bibnamefont
  {Bertoldi}}, \bibinfo {author} {\bibfnamefont {F.}~\bibnamefont {Minardi}},\
  and\ \bibinfo {author} {\bibfnamefont {M.}~\bibnamefont {Prevedelli}},\
  }\bibfield  {title} {\bibinfo {title} {Phase shift in atom interferometers:
  Corrections for nonquadratic potentials and finite-duration laser pulses},\
  }\href {https://doi.org/10.1103/PhysRevA.99.033619} {\bibfield  {journal}
  {\bibinfo  {journal} {Phys. Rev. A}\ }\textbf {\bibinfo {volume} {99}},\
  \bibinfo {pages} {033619} (\bibinfo {year} {2019})}\BibitemShut {NoStop}%
\bibitem [{\citenamefont {Asenbaum}\ \emph {et~al.}(2017)\citenamefont
  {Asenbaum}, \citenamefont {Overstreet}, \citenamefont {Kovachy},
  \citenamefont {Brown}, \citenamefont {Hogan},\ and\ \citenamefont
  {Kasevich}}]{PhysRevLett.118.183602}%
  \BibitemOpen
  \bibfield  {author} {\bibinfo {author} {\bibfnamefont {P.}~\bibnamefont
  {Asenbaum}}, \bibinfo {author} {\bibfnamefont {C.}~\bibnamefont
  {Overstreet}}, \bibinfo {author} {\bibfnamefont {T.}~\bibnamefont {Kovachy}},
  \bibinfo {author} {\bibfnamefont {D.~D.}\ \bibnamefont {Brown}}, \bibinfo
  {author} {\bibfnamefont {J.~M.}\ \bibnamefont {Hogan}},\ and\ \bibinfo
  {author} {\bibfnamefont {M.~A.}\ \bibnamefont {Kasevich}},\ }\bibfield
  {title} {\bibinfo {title} {Phase shift in an atom interferometer due to
  spacetime curvature across its wave function},\ }\href
  {https://doi.org/10.1103/PhysRevLett.118.183602} {\bibfield  {journal}
  {\bibinfo  {journal} {Phys. Rev. Lett.}\ }\textbf {\bibinfo {volume} {118}},\
  \bibinfo {pages} {183602} (\bibinfo {year} {2017})}\BibitemShut {NoStop}%
\bibitem [{\citenamefont {Cleland}\ and\ \citenamefont
  {Roukes}(1998)}]{Cleland1998}%
  \BibitemOpen
  \bibfield  {author} {\bibinfo {author} {\bibfnamefont {A.~N.}\ \bibnamefont
  {Cleland}}\ and\ \bibinfo {author} {\bibfnamefont {M.~L.}\ \bibnamefont
  {Roukes}},\ }\bibfield  {title} {\bibinfo {title} {A nanometre-scale
  mechanical electrometer},\ }\href {https://doi.org/10.1038/32373} {\bibfield
  {journal} {\bibinfo  {journal} {Nature}\ }\textbf {\bibinfo {volume} {392}},\
  \bibinfo {pages} {160} (\bibinfo {year} {1998})}\BibitemShut {NoStop}%
\bibitem [{\citenamefont {Schneider}\ \emph {et~al.}(2017)\citenamefont
  {Schneider}, \citenamefont {Mooser}, \citenamefont {Bohman}, \citenamefont
  {Sch{\"o}n}, \citenamefont {Harrington}, \citenamefont {Higuchi},
  \citenamefont {Nagahama}, \citenamefont {Sellner}, \citenamefont {Smorra},
  \citenamefont {Blaum}, \citenamefont {Matsuda}, \citenamefont {Quint},
  \citenamefont {Walz},\ and\ \citenamefont {Ulmer}}]{Schneider1081}%
  \BibitemOpen
  \bibfield  {author} {\bibinfo {author} {\bibfnamefont {G.}~\bibnamefont
  {Schneider}}, \bibinfo {author} {\bibfnamefont {A.}~\bibnamefont {Mooser}},
  \bibinfo {author} {\bibfnamefont {M.}~\bibnamefont {Bohman}}, \bibinfo
  {author} {\bibfnamefont {N.}~\bibnamefont {Sch{\"o}n}}, \bibinfo {author}
  {\bibfnamefont {J.}~\bibnamefont {Harrington}}, \bibinfo {author}
  {\bibfnamefont {T.}~\bibnamefont {Higuchi}}, \bibinfo {author} {\bibfnamefont
  {H.}~\bibnamefont {Nagahama}}, \bibinfo {author} {\bibfnamefont
  {S.}~\bibnamefont {Sellner}}, \bibinfo {author} {\bibfnamefont
  {C.}~\bibnamefont {Smorra}}, \bibinfo {author} {\bibfnamefont
  {K.}~\bibnamefont {Blaum}}, \bibinfo {author} {\bibfnamefont
  {Y.}~\bibnamefont {Matsuda}}, \bibinfo {author} {\bibfnamefont
  {W.}~\bibnamefont {Quint}}, \bibinfo {author} {\bibfnamefont
  {J.}~\bibnamefont {Walz}},\ and\ \bibinfo {author} {\bibfnamefont
  {S.}~\bibnamefont {Ulmer}},\ }\bibfield  {title} {\bibinfo {title}
  {Double-trap measurement of the proton magnetic moment at 0.3 parts per
  billion precision},\ }\href {https://doi.org/10.1126/science.aan0207}
  {\bibfield  {journal} {\bibinfo  {journal} {Science}\ }\textbf {\bibinfo
  {volume} {358}},\ \bibinfo {pages} {1081} (\bibinfo {year} {2017})},\ \Eprint
  {https://arxiv.org/abs/https://science.sciencemag.org/content/358/6366/1081.full.pdf}
  {https://science.sciencemag.org/content/358/6366/1081.full.pdf} \BibitemShut
  {NoStop}%
\bibitem [{\citenamefont {Harber}\ \emph {et~al.}(2005)\citenamefont {Harber},
  \citenamefont {Obrecht}, \citenamefont {McGuirk},\ and\ \citenamefont
  {Cornell}}]{PhysRevA.72.033610}%
  \BibitemOpen
  \bibfield  {author} {\bibinfo {author} {\bibfnamefont {D.~M.}\ \bibnamefont
  {Harber}}, \bibinfo {author} {\bibfnamefont {J.~M.}\ \bibnamefont {Obrecht}},
  \bibinfo {author} {\bibfnamefont {J.~M.}\ \bibnamefont {McGuirk}},\ and\
  \bibinfo {author} {\bibfnamefont {E.~A.}\ \bibnamefont {Cornell}},\
  }\bibfield  {title} {\bibinfo {title} {Measurement of the casimir-polder
  force through center-of-mass oscillations of a bose-einstein condensate},\
  }\href {https://doi.org/10.1103/PhysRevA.72.033610} {\bibfield  {journal}
  {\bibinfo  {journal} {Phys. Rev. A}\ }\textbf {\bibinfo {volume} {72}},\
  \bibinfo {pages} {033610} (\bibinfo {year} {2005})}\BibitemShut {NoStop}%
\bibitem [{\citenamefont {Obrecht}\ \emph {et~al.}(2007)\citenamefont
  {Obrecht}, \citenamefont {Wild}, \citenamefont {Antezza}, \citenamefont
  {Pitaevskii}, \citenamefont {Stringari},\ and\ \citenamefont
  {Cornell}}]{PhysRevLett.98.063201}%
  \BibitemOpen
  \bibfield  {author} {\bibinfo {author} {\bibfnamefont {J.~M.}\ \bibnamefont
  {Obrecht}}, \bibinfo {author} {\bibfnamefont {R.~J.}\ \bibnamefont {Wild}},
  \bibinfo {author} {\bibfnamefont {M.}~\bibnamefont {Antezza}}, \bibinfo
  {author} {\bibfnamefont {L.~P.}\ \bibnamefont {Pitaevskii}}, \bibinfo
  {author} {\bibfnamefont {S.}~\bibnamefont {Stringari}},\ and\ \bibinfo
  {author} {\bibfnamefont {E.~A.}\ \bibnamefont {Cornell}},\ }\bibfield
  {title} {\bibinfo {title} {Measurement of the temperature dependence of the
  casimir-polder force},\ }\href
  {https://doi.org/10.1103/PhysRevLett.98.063201} {\bibfield  {journal}
  {\bibinfo  {journal} {Phys. Rev. Lett.}\ }\textbf {\bibinfo {volume} {98}},\
  \bibinfo {pages} {063201} (\bibinfo {year} {2007})}\BibitemShut {NoStop}%
\bibitem [{\citenamefont {Chang}\ \emph {et~al.}(2016)\citenamefont {Chang},
  \citenamefont {Bouton}, \citenamefont {Cayla}, \citenamefont {Qu},
  \citenamefont {Aspect}, \citenamefont {Westbrook},\ and\ \citenamefont
  {Cl\'ement}}]{PhysRevLett.117.235303}%
  \BibitemOpen
  \bibfield  {author} {\bibinfo {author} {\bibfnamefont {R.}~\bibnamefont
  {Chang}}, \bibinfo {author} {\bibfnamefont {Q.}~\bibnamefont {Bouton}},
  \bibinfo {author} {\bibfnamefont {H.}~\bibnamefont {Cayla}}, \bibinfo
  {author} {\bibfnamefont {C.}~\bibnamefont {Qu}}, \bibinfo {author}
  {\bibfnamefont {A.}~\bibnamefont {Aspect}}, \bibinfo {author} {\bibfnamefont
  {C.~I.}\ \bibnamefont {Westbrook}},\ and\ \bibinfo {author} {\bibfnamefont
  {D.}~\bibnamefont {Cl\'ement}},\ }\bibfield  {title} {\bibinfo {title}
  {Momentum-resolved observation of thermal and quantum depletion in a bose
  gas},\ }\href {https://doi.org/10.1103/PhysRevLett.117.235303} {\bibfield
  {journal} {\bibinfo  {journal} {Phys. Rev. Lett.}\ }\textbf {\bibinfo
  {volume} {117}},\ \bibinfo {pages} {235303} (\bibinfo {year}
  {2016})}\BibitemShut {NoStop}%
\bibitem [{\citenamefont {{Ross}}\ \emph {et~al.}(2021)\citenamefont {{Ross}},
  \citenamefont {{Deuar}}, \citenamefont {{Shin}}, \citenamefont {{Thomas}},
  \citenamefont {{Henson}}, \citenamefont {{Hodgman}},\ and\ \citenamefont
  {{Truscott}}}]{rossQD}%
  \BibitemOpen
  \bibfield  {author} {\bibinfo {author} {\bibfnamefont {J.~A.}\ \bibnamefont
  {{Ross}}}, \bibinfo {author} {\bibfnamefont {P.}~\bibnamefont {{Deuar}}},
  \bibinfo {author} {\bibfnamefont {D.~K.}\ \bibnamefont {{Shin}}}, \bibinfo
  {author} {\bibfnamefont {K.~F.}\ \bibnamefont {{Thomas}}}, \bibinfo {author}
  {\bibfnamefont {B.~M.}\ \bibnamefont {{Henson}}}, \bibinfo {author}
  {\bibfnamefont {S.~S.}\ \bibnamefont {{Hodgman}}},\ and\ \bibinfo {author}
  {\bibfnamefont {A.~G.}\ \bibnamefont {{Truscott}}},\ }\bibfield  {title}
  {\bibinfo {title} {{Survival of the quantum depletion of a Bose-Einstein
  condensate after release from a magnetic trap}},\ }\href@noop {} {\bibfield
  {journal} {\bibinfo  {journal} {arXiv e-prints}\ ,\ \bibinfo {eid}
  {arXiv:2103.15283}} (\bibinfo {year} {2021})},\ \Eprint
  {https://arxiv.org/abs/2103.15283} {arXiv:2103.15283 [cond-mat.quant-gas]}
  \BibitemShut {NoStop}%
\bibitem [{\citenamefont {Tenart}\ \emph {et~al.}(2021)\citenamefont {Tenart},
  \citenamefont {Herc{\'{e}}}, \citenamefont {Bureik}, \citenamefont {Dareau},\
  and\ \citenamefont {Cl{\'{e}}ment}}]{Tenart2021}%
  \BibitemOpen
  \bibfield  {author} {\bibinfo {author} {\bibfnamefont {A.}~\bibnamefont
  {Tenart}}, \bibinfo {author} {\bibfnamefont {G.}~\bibnamefont {Herc{\'{e}}}},
  \bibinfo {author} {\bibfnamefont {J.-P.}\ \bibnamefont {Bureik}}, \bibinfo
  {author} {\bibfnamefont {A.}~\bibnamefont {Dareau}},\ and\ \bibinfo {author}
  {\bibfnamefont {D.}~\bibnamefont {Cl{\'{e}}ment}},\ }\bibfield  {title}
  {\bibinfo {title} {Observation of pairs of atoms at opposite momenta in an
  equilibrium interacting bose gas},\ }\href
  {https://doi.org/10.1038/s41567-021-01381-2} {\bibfield  {journal} {\bibinfo
  {journal} {Nature Physics}\ }\textbf {\bibinfo {volume} {17}},\ \bibinfo
  {pages} {1364} (\bibinfo {year} {2021})}\BibitemShut {NoStop}%
\bibitem [{\citenamefont {Brown}\ \emph {et~al.}(2018)\citenamefont {Brown},
  \citenamefont {McPhail}, \citenamefont {White}, \citenamefont {Baillie},
  \citenamefont {Ruddell},\ and\ \citenamefont {Hoogerland}}]{Brown2018}%
  \BibitemOpen
  \bibfield  {author} {\bibinfo {author} {\bibfnamefont {D.~J.}\ \bibnamefont
  {Brown}}, \bibinfo {author} {\bibfnamefont {A.~V.~H.}\ \bibnamefont
  {McPhail}}, \bibinfo {author} {\bibfnamefont {D.~H.}\ \bibnamefont {White}},
  \bibinfo {author} {\bibfnamefont {D.}~\bibnamefont {Baillie}}, \bibinfo
  {author} {\bibfnamefont {S.~K.}\ \bibnamefont {Ruddell}},\ and\ \bibinfo
  {author} {\bibfnamefont {M.~D.}\ \bibnamefont {Hoogerland}},\ }\bibfield
  {title} {\bibinfo {title} {Thermalization, condensate growth, and defect
  formation in an out-of-equilibrium bose gas},\ }\bibfield  {journal}
  {\bibinfo  {journal} {Physical Review A}\ }\textbf {\bibinfo {volume} {98}},\
  \href {https://doi.org/10.1103/physreva.98.013606}
  {10.1103/physreva.98.013606} (\bibinfo {year} {2018})\BibitemShut {NoStop}%
\bibitem [{\citenamefont {Bartenstein}\ \emph {et~al.}(2004)\citenamefont
  {Bartenstein}, \citenamefont {Altmeyer}, \citenamefont {Riedl}, \citenamefont
  {Jochim}, \citenamefont {Chin}, \citenamefont {Denschlag},\ and\
  \citenamefont {Grimm}}]{Bartenstein2004}%
  \BibitemOpen
  \bibfield  {author} {\bibinfo {author} {\bibfnamefont {M.}~\bibnamefont
  {Bartenstein}}, \bibinfo {author} {\bibfnamefont {A.}~\bibnamefont
  {Altmeyer}}, \bibinfo {author} {\bibfnamefont {S.}~\bibnamefont {Riedl}},
  \bibinfo {author} {\bibfnamefont {S.}~\bibnamefont {Jochim}}, \bibinfo
  {author} {\bibfnamefont {C.}~\bibnamefont {Chin}}, \bibinfo {author}
  {\bibfnamefont {J.~H.}\ \bibnamefont {Denschlag}},\ and\ \bibinfo {author}
  {\bibfnamefont {R.}~\bibnamefont {Grimm}},\ }\bibfield  {title} {\bibinfo
  {title} {Collective excitations of a degenerate gas at the {BEC}-{BCS}
  crossover},\ }\bibfield  {journal} {\bibinfo  {journal} {Physical Review
  Letters}\ }\textbf {\bibinfo {volume} {92}},\ \href
  {https://doi.org/10.1103/physrevlett.92.203201}
  {10.1103/physrevlett.92.203201} (\bibinfo {year} {2004})\BibitemShut
  {NoStop}%
\bibitem [{\citenamefont {Altmeyer}\ \emph {et~al.}(2007)\citenamefont
  {Altmeyer}, \citenamefont {Riedl}, \citenamefont {Kohstall}, \citenamefont
  {Wright}, \citenamefont {Geursen}, \citenamefont {Bartenstein}, \citenamefont
  {Chin}, \citenamefont {Denschlag},\ and\ \citenamefont
  {Grimm}}]{PhysRevLett.98.040401}%
  \BibitemOpen
  \bibfield  {author} {\bibinfo {author} {\bibfnamefont {A.}~\bibnamefont
  {Altmeyer}}, \bibinfo {author} {\bibfnamefont {S.}~\bibnamefont {Riedl}},
  \bibinfo {author} {\bibfnamefont {C.}~\bibnamefont {Kohstall}}, \bibinfo
  {author} {\bibfnamefont {M.~J.}\ \bibnamefont {Wright}}, \bibinfo {author}
  {\bibfnamefont {R.}~\bibnamefont {Geursen}}, \bibinfo {author} {\bibfnamefont
  {M.}~\bibnamefont {Bartenstein}}, \bibinfo {author} {\bibfnamefont
  {C.}~\bibnamefont {Chin}}, \bibinfo {author} {\bibfnamefont {J.~H.}\
  \bibnamefont {Denschlag}},\ and\ \bibinfo {author} {\bibfnamefont
  {R.}~\bibnamefont {Grimm}},\ }\bibfield  {title} {\bibinfo {title} {Precision
  measurements of collective oscillations in the bec-bcs crossover},\ }\href
  {https://doi.org/10.1103/PhysRevLett.98.040401} {\bibfield  {journal}
  {\bibinfo  {journal} {Phys. Rev. Lett.}\ }\textbf {\bibinfo {volume} {98}},\
  \bibinfo {pages} {040401} (\bibinfo {year} {2007})}\BibitemShut {NoStop}%
\bibitem [{\citenamefont {Nagler}\ \emph {et~al.}(2020)\citenamefont {Nagler},
  \citenamefont {J\"agering}, \citenamefont {Sheikhan}, \citenamefont
  {Barbosa}, \citenamefont {Koch}, \citenamefont {Eggert}, \citenamefont
  {Schneider},\ and\ \citenamefont {Widera}}]{PhysRevA.101.053633}%
  \BibitemOpen
  \bibfield  {author} {\bibinfo {author} {\bibfnamefont {B.}~\bibnamefont
  {Nagler}}, \bibinfo {author} {\bibfnamefont {K.}~\bibnamefont {J\"agering}},
  \bibinfo {author} {\bibfnamefont {A.}~\bibnamefont {Sheikhan}}, \bibinfo
  {author} {\bibfnamefont {S.}~\bibnamefont {Barbosa}}, \bibinfo {author}
  {\bibfnamefont {J.}~\bibnamefont {Koch}}, \bibinfo {author} {\bibfnamefont
  {S.}~\bibnamefont {Eggert}}, \bibinfo {author} {\bibfnamefont
  {I.}~\bibnamefont {Schneider}},\ and\ \bibinfo {author} {\bibfnamefont
  {A.}~\bibnamefont {Widera}},\ }\bibfield  {title} {\bibinfo {title} {Dipole
  oscillations of fermionic quantum gases along the bec-bcs crossover in
  disordered potentials},\ }\href {https://doi.org/10.1103/PhysRevA.101.053633}
  {\bibfield  {journal} {\bibinfo  {journal} {Phys. Rev. A}\ }\textbf {\bibinfo
  {volume} {101}},\ \bibinfo {pages} {053633} (\bibinfo {year}
  {2020})}\BibitemShut {NoStop}%
\bibitem [{\citenamefont {Henson}\ \emph {et~al.}(2021)\citenamefont {Henson},
  \citenamefont {Ross}, \citenamefont {Thomas}, \citenamefont {Kuhn},
  \citenamefont {Shin}, \citenamefont {Hodgman}, \citenamefont {Zhang},
  \citenamefont {Tang}, \citenamefont {Drake}, \citenamefont {Bondy},
  \citenamefont {Truscott},\ and\ \citenamefont
  {Baldwin}}]{henson2021precision}%
  \BibitemOpen
  \bibfield  {author} {\bibinfo {author} {\bibfnamefont {B.~M.}\ \bibnamefont
  {Henson}}, \bibinfo {author} {\bibfnamefont {J.~A.}\ \bibnamefont {Ross}},
  \bibinfo {author} {\bibfnamefont {K.~F.}\ \bibnamefont {Thomas}}, \bibinfo
  {author} {\bibfnamefont {C.~N.}\ \bibnamefont {Kuhn}}, \bibinfo {author}
  {\bibfnamefont {D.~K.}\ \bibnamefont {Shin}}, \bibinfo {author}
  {\bibfnamefont {S.~S.}\ \bibnamefont {Hodgman}}, \bibinfo {author}
  {\bibfnamefont {Y.-H.}\ \bibnamefont {Zhang}}, \bibinfo {author}
  {\bibfnamefont {L.-Y.}\ \bibnamefont {Tang}}, \bibinfo {author}
  {\bibfnamefont {G.~W.~F.}\ \bibnamefont {Drake}}, \bibinfo {author}
  {\bibfnamefont {A.~T.}\ \bibnamefont {Bondy}}, \bibinfo {author}
  {\bibfnamefont {A.~G.}\ \bibnamefont {Truscott}},\ and\ \bibinfo {author}
  {\bibfnamefont {K.~G.~H.}\ \bibnamefont {Baldwin}},\ }\href@noop {} {\bibinfo
  {title} {Precision measurement of the helium $2^{3\!}s_1- 2^{3\!}p/3^{3\!}p$
  tune-out frequency as a test of qed}} (\bibinfo {year} {2021}),\ \Eprint
  {https://arxiv.org/abs/2107.00149} {arXiv:2107.00149 [physics.atom-ph]}
  \BibitemShut {NoStop}%
\bibitem [{\citenamefont {Manning}\ \emph {et~al.}(2010)\citenamefont
  {Manning}, \citenamefont {Hodgman}, \citenamefont {Dall}, \citenamefont
  {Johnsson},\ and\ \citenamefont {Truscott}}]{Manning:10}%
  \BibitemOpen
  \bibfield  {author} {\bibinfo {author} {\bibfnamefont {A.~G.}\ \bibnamefont
  {Manning}}, \bibinfo {author} {\bibfnamefont {S.~S.}\ \bibnamefont
  {Hodgman}}, \bibinfo {author} {\bibfnamefont {R.~G.}\ \bibnamefont {Dall}},
  \bibinfo {author} {\bibfnamefont {M.~T.}\ \bibnamefont {Johnsson}},\ and\
  \bibinfo {author} {\bibfnamefont {A.~G.}\ \bibnamefont {Truscott}},\
  }\bibfield  {title} {\bibinfo {title} {The hanbury brown-twiss effect in a
  pulsed atom laser},\ }\href {https://doi.org/10.1364/OE.18.018712} {\bibfield
   {journal} {\bibinfo  {journal} {Opt. Express}\ }\textbf {\bibinfo {volume}
  {18}},\ \bibinfo {pages} {18712} (\bibinfo {year} {2010})}\BibitemShut
  {NoStop}%
\bibitem [{\citenamefont {Henson}\ \emph
  {et~al.}(2018{\natexlab{a}})\citenamefont {Henson}, \citenamefont {Shin},
  \citenamefont {Thomas}, \citenamefont {Ross}, \citenamefont {Hush},
  \citenamefont {Hodgman},\ and\ \citenamefont {Truscott}}]{Henson13216}%
  \BibitemOpen
  \bibfield  {author} {\bibinfo {author} {\bibfnamefont {B.~M.}\ \bibnamefont
  {Henson}}, \bibinfo {author} {\bibfnamefont {D.~K.}\ \bibnamefont {Shin}},
  \bibinfo {author} {\bibfnamefont {K.~F.}\ \bibnamefont {Thomas}}, \bibinfo
  {author} {\bibfnamefont {J.~A.}\ \bibnamefont {Ross}}, \bibinfo {author}
  {\bibfnamefont {M.~R.}\ \bibnamefont {Hush}}, \bibinfo {author}
  {\bibfnamefont {S.~S.}\ \bibnamefont {Hodgman}},\ and\ \bibinfo {author}
  {\bibfnamefont {A.~G.}\ \bibnamefont {Truscott}},\ }\bibfield  {title}
  {\bibinfo {title} {Approaching the adiabatic timescale with machine
  learning},\ }\href {https://doi.org/10.1073/pnas.1811501115} {\bibfield
  {journal} {\bibinfo  {journal} {Proceedings of the National Academy of
  Sciences}\ }\textbf {\bibinfo {volume} {115}},\ \bibinfo {pages} {13216}
  (\bibinfo {year} {2018}{\natexlab{a}})},\ \Eprint
  {https://arxiv.org/abs/https://www.pnas.org/content/115/52/13216.full.pdf}
  {https://www.pnas.org/content/115/52/13216.full.pdf} \BibitemShut {NoStop}%
\bibitem [{\citenamefont {Smirne}(2005)}]{smirne2005experiments}%
  \BibitemOpen
  \bibfield  {author} {\bibinfo {author} {\bibfnamefont {G.}~\bibnamefont
  {Smirne}},\ }\emph {\bibinfo {title} {Experiments with Bose-Einstein
  condensates in optical traps}},\ \href
  {https://www2.physics.ox.ac.uk/sites/default/files/2013-01-19/giuseppe_pdf_75967.pdf}
  {Ph.D. thesis},\ \bibinfo  {school} {University of Oxford} (\bibinfo {year}
  {2005})\BibitemShut {NoStop}%
\bibitem [{\citenamefont {Humbert}(2012)}]{LHumbertPHD}%
  \BibitemOpen
  \bibfield  {author} {\bibinfo {author} {\bibfnamefont {L.~H.}\ \bibnamefont
  {Humbert}},\ }\emph {\bibinfo {title} {All-optical Rb87 Bose-Einstein
  condensate apparatus: construction and operation}},\ \href
  {https://bec.equs.org/sites/default/files/content/s41320046_phd_finalthesis.pdf}
  {Ph.D. thesis},\ \bibinfo  {school} {The University of Queensland} (\bibinfo
  {year} {2012})\BibitemShut {NoStop}%
\bibitem [{\citenamefont {Wang}(2006)}]{wangphd}%
  \BibitemOpen
  \bibfield  {author} {\bibinfo {author} {\bibfnamefont {R.}~\bibnamefont
  {Wang}},\ }\emph {\bibinfo {title} {Approaching lithium BEC with a mini
  trap}},\ \href
  {https://web.stanford.edu/group/kasevich/cgi-bin/wordpress/wp-content/uploads/2012/09/WangThesis.pdf}
  {Ph.D. thesis},\ \bibinfo  {school} {Yale University} (\bibinfo {year}
  {2006})\BibitemShut {NoStop}%
\bibitem [{\citenamefont {He}\ \emph {et~al.}(2012)\citenamefont {He},
  \citenamefont {Yu}, \citenamefont {Xu}, \citenamefont {Wang},\ and\
  \citenamefont {Zhan}}]{OE.20.003711}%
  \BibitemOpen
  \bibfield  {author} {\bibinfo {author} {\bibfnamefont {X.}~\bibnamefont
  {He}}, \bibinfo {author} {\bibfnamefont {S.}~\bibnamefont {Yu}}, \bibinfo
  {author} {\bibfnamefont {P.}~\bibnamefont {Xu}}, \bibinfo {author}
  {\bibfnamefont {J.}~\bibnamefont {Wang}},\ and\ \bibinfo {author}
  {\bibfnamefont {M.}~\bibnamefont {Zhan}},\ }\bibfield  {title} {\bibinfo
  {title} {Combining red and blue-detuned optical potentials to form a
  lamb-dicke trap for a single neutral atom},\ }\href
  {https://doi.org/10.1364/OE.20.003711} {\bibfield  {journal} {\bibinfo
  {journal} {Opt. Express}\ }\textbf {\bibinfo {volume} {20}},\ \bibinfo
  {pages} {3711} (\bibinfo {year} {2012})}\BibitemShut {NoStop}%
\bibitem [{\citenamefont {Makhalov}\ \emph {et~al.}(2015)\citenamefont
  {Makhalov}, \citenamefont {Martiyanov}, \citenamefont {Barmashova},\ and\
  \citenamefont {Turlapov}}]{MAKHALOV2015327}%
  \BibitemOpen
  \bibfield  {author} {\bibinfo {author} {\bibfnamefont {V.}~\bibnamefont
  {Makhalov}}, \bibinfo {author} {\bibfnamefont {K.}~\bibnamefont
  {Martiyanov}}, \bibinfo {author} {\bibfnamefont {T.}~\bibnamefont
  {Barmashova}},\ and\ \bibinfo {author} {\bibfnamefont {A.}~\bibnamefont
  {Turlapov}},\ }\bibfield  {title} {\bibinfo {title} {Precision measurement of
  a trapping potential for an ultracold gas},\ }\href
  {https://doi.org/https://doi.org/10.1016/j.physleta.2014.10.049} {\bibfield
  {journal} {\bibinfo  {journal} {Physics Letters A}\ }\textbf {\bibinfo
  {volume} {379}},\ \bibinfo {pages} {327 } (\bibinfo {year}
  {2015})}\BibitemShut {NoStop}%
\bibitem [{\citenamefont {Llorente~Garc\'{\i}a}\ \emph
  {et~al.}(2013)\citenamefont {Llorente~Garc\'{\i}a}, \citenamefont
  {Darqui\'e}, \citenamefont {Sinclair}, \citenamefont {Curtis}, \citenamefont
  {Tachikawa}, \citenamefont {Hudson},\ and\ \citenamefont
  {Hinds}}]{PhysRevA.88.043406}%
  \BibitemOpen
  \bibfield  {author} {\bibinfo {author} {\bibfnamefont {I.}~\bibnamefont
  {Llorente~Garc\'{\i}a}}, \bibinfo {author} {\bibfnamefont {B.}~\bibnamefont
  {Darqui\'e}}, \bibinfo {author} {\bibfnamefont {C.~D.~J.}\ \bibnamefont
  {Sinclair}}, \bibinfo {author} {\bibfnamefont {E.~A.}\ \bibnamefont
  {Curtis}}, \bibinfo {author} {\bibfnamefont {M.}~\bibnamefont {Tachikawa}},
  \bibinfo {author} {\bibfnamefont {J.~J.}\ \bibnamefont {Hudson}},\ and\
  \bibinfo {author} {\bibfnamefont {E.~A.}\ \bibnamefont {Hinds}},\ }\bibfield
  {title} {\bibinfo {title} {Shaking-induced dynamics of cold atoms in magnetic
  traps},\ }\href {https://doi.org/10.1103/PhysRevA.88.043406} {\bibfield
  {journal} {\bibinfo  {journal} {Phys. Rev. A}\ }\textbf {\bibinfo {volume}
  {88}},\ \bibinfo {pages} {043406} (\bibinfo {year} {2013})}\BibitemShut
  {NoStop}%
\bibitem [{\citenamefont {Wang}(2017)}]{yibophd}%
  \BibitemOpen
  \bibfield  {author} {\bibinfo {author} {\bibfnamefont {Y.}~\bibnamefont
  {Wang}},\ }\emph {\bibinfo {title} {Sub-Micron-Period Magnetic Lattices for
  Ultracold Atoms}},\ \href
  {https://www.swinburne.edu.au/engineering/caous/theses/Yibo%20thesis.pdf}
  {Ph.D. thesis},\ \bibinfo  {school} {Swinburne University of Technology}
  (\bibinfo {year} {2017})\BibitemShut {NoStop}%
\bibitem [{\citenamefont {Whitlock}(2007)}]{Whitlockphd}%
  \BibitemOpen
  \bibfield  {author} {\bibinfo {author} {\bibfnamefont {S.}~\bibnamefont
  {Whitlock}},\ }\emph {\bibinfo {title} {Bose-Einstein condensates on a
  magnetic film atom chip}},\ \href
  {http://www.swinburne.edu.au/engineering/caous/theses/Whitlock\%20Thesis.pdf}
  {Ph.D. thesis},\ \bibinfo  {school} {Swinburne University of Technology}
  (\bibinfo {year} {2007})\BibitemShut {NoStop}%
\bibitem [{\citenamefont {Altmeyer}(2007)}]{Altmeyerphd}%
  \BibitemOpen
  \bibfield  {author} {\bibinfo {author} {\bibfnamefont {A.}~\bibnamefont
  {Altmeyer}},\ }\emph {\bibinfo {title} {Collective oscillationsof an
  ultracold quantum gasin the BEC-BCS crossover regime}},\ \href
  {http://www.ultracold.at/theses/2007-altmeyer.pdf} {Ph.D. thesis},\ \bibinfo
  {school} {University of Innsbruck} (\bibinfo {year} {2007})\BibitemShut
  {NoStop}%
\bibitem [{\citenamefont {Berrada}(2016)}]{berradaphd}%
  \BibitemOpen
  \bibfield  {author} {\bibinfo {author} {\bibfnamefont {T.}~\bibnamefont
  {Berrada}},\ }\emph {\bibinfo {title} {Interferometry with Interacting
  Bose-Einstein Condensates in a Double-Well Potential}},\ \href
  {dx.do.org/10.1007/978-3-319-27233-7} {Ph.D. thesis},\ \bibinfo  {school}
  {Vienna University of Technology} (\bibinfo {year} {2016})\BibitemShut
  {NoStop}%
\bibitem [{\citenamefont {Theis}(2005)}]{MatthiasTheisphd}%
  \BibitemOpen
  \bibfield  {author} {\bibinfo {author} {\bibfnamefont {M.}~\bibnamefont
  {Theis}},\ }\emph {\bibinfo {title} {Optical Feshbach Resonancesin a
  Bose-Einstein Condensate}},\ \href
  {http://www.ultracold.at/theses/2005-theis.pdf} {Ph.D. thesis},\ \bibinfo
  {school} {Leopold-Franzens-Universit\"at Innsbruck} (\bibinfo {year}
  {2005})\BibitemShut {NoStop}%
\bibitem [{\citenamefont {Gajdacz}\ \emph {et~al.}(2013)\citenamefont
  {Gajdacz}, \citenamefont {Pedersen}, \citenamefont {Mørch}, \citenamefont
  {Hilliard}, \citenamefont {Arlt},\ and\ \citenamefont
  {Sherson}}]{doi:10.1063/1.4818913}%
  \BibitemOpen
  \bibfield  {author} {\bibinfo {author} {\bibfnamefont {M.}~\bibnamefont
  {Gajdacz}}, \bibinfo {author} {\bibfnamefont {P.~L.}\ \bibnamefont
  {Pedersen}}, \bibinfo {author} {\bibfnamefont {T.}~\bibnamefont {Mørch}},
  \bibinfo {author} {\bibfnamefont {A.~J.}\ \bibnamefont {Hilliard}}, \bibinfo
  {author} {\bibfnamefont {J.}~\bibnamefont {Arlt}},\ and\ \bibinfo {author}
  {\bibfnamefont {J.~F.}\ \bibnamefont {Sherson}},\ }\bibfield  {title}
  {\bibinfo {title} {Non-destructive faraday imaging of dynamically controlled
  ultracold atoms},\ }\href {https://doi.org/10.1063/1.4818913} {\bibfield
  {journal} {\bibinfo  {journal} {Review of Scientific Instruments}\ }\textbf
  {\bibinfo {volume} {84}},\ \bibinfo {pages} {083105} (\bibinfo {year}
  {2013})},\ \Eprint {https://arxiv.org/abs/https://doi.org/10.1063/1.4818913}
  {https://doi.org/10.1063/1.4818913} \BibitemShut {NoStop}%
\bibitem [{\citenamefont {Petrov}\ \emph {et~al.}(2007)\citenamefont {Petrov},
  \citenamefont {Oblak}, \citenamefont {Alzar}, \citenamefont {Kj\ae{}rgaard},\
  and\ \citenamefont {Polzik}}]{PhysRevA.75.033803}%
  \BibitemOpen
  \bibfield  {author} {\bibinfo {author} {\bibfnamefont {P.~G.}\ \bibnamefont
  {Petrov}}, \bibinfo {author} {\bibfnamefont {D.}~\bibnamefont {Oblak}},
  \bibinfo {author} {\bibfnamefont {C.~L.~G.}\ \bibnamefont {Alzar}}, \bibinfo
  {author} {\bibfnamefont {N.}~\bibnamefont {Kj\ae{}rgaard}},\ and\ \bibinfo
  {author} {\bibfnamefont {E.~S.}\ \bibnamefont {Polzik}},\ }\bibfield  {title}
  {\bibinfo {title} {Nondestructive interferometric characterization of an
  optical dipole trap},\ }\href {https://doi.org/10.1103/PhysRevA.75.033803}
  {\bibfield  {journal} {\bibinfo  {journal} {Phys. Rev. A}\ }\textbf {\bibinfo
  {volume} {75}},\ \bibinfo {pages} {033803} (\bibinfo {year}
  {2007})}\BibitemShut {NoStop}%
\bibitem [{\citenamefont {Oblak}\ \emph {et~al.}(2015)\citenamefont {Oblak},
  \citenamefont {Appel}, \citenamefont {Windpassinger}, \citenamefont {Hoff},
  \citenamefont {Kj{\ae}rgaard},\ and\ \citenamefont {Polzik}}]{Oblak2015}%
  \BibitemOpen
  \bibfield  {author} {\bibinfo {author} {\bibfnamefont {D.}~\bibnamefont
  {Oblak}}, \bibinfo {author} {\bibfnamefont {J.}~\bibnamefont {Appel}},
  \bibinfo {author} {\bibfnamefont {P.}~\bibnamefont {Windpassinger}}, \bibinfo
  {author} {\bibfnamefont {U.}~\bibnamefont {Hoff}}, \bibinfo {author}
  {\bibfnamefont {N.}~\bibnamefont {Kj{\ae}rgaard}},\ and\ \bibinfo {author}
  {\bibfnamefont {E.}~\bibnamefont {Polzik}},\ }\bibfield  {title} {\bibinfo
  {title} {Echo spectroscopy of atomic dynamics in a gaussian trap via phase
  imprints},\ }\href {https://doi.org/10.1140/epjd/e2008-00192-1} {\bibfield
  {journal} {\bibinfo  {journal} {The European Physical Journal D: Atomic,
  Molecular, Optical and Plasma Physics}\ }\textbf {\bibinfo {volume} {50}},\
  \bibinfo {pages} {67–73} (\bibinfo {year} {2015})}\BibitemShut {NoStop}%
\bibitem [{\citenamefont {Kohnen}\ \emph {et~al.}(2011)\citenamefont {Kohnen},
  \citenamefont {Petrov}, \citenamefont {Nyman},\ and\ \citenamefont
  {Hinds}}]{Kohnen_2011}%
  \BibitemOpen
  \bibfield  {author} {\bibinfo {author} {\bibfnamefont {M.}~\bibnamefont
  {Kohnen}}, \bibinfo {author} {\bibfnamefont {P.~G.}\ \bibnamefont {Petrov}},
  \bibinfo {author} {\bibfnamefont {R.~A.}\ \bibnamefont {Nyman}},\ and\
  \bibinfo {author} {\bibfnamefont {E.~A.}\ \bibnamefont {Hinds}},\ }\bibfield
  {title} {\bibinfo {title} {Minimally destructive detection of magnetically
  trapped atoms using frequency-synthesized light},\ }\href
  {https://doi.org/10.1088/1367-2630/13/8/085006} {\bibfield  {journal}
  {\bibinfo  {journal} {New Journal of Physics}\ }\textbf {\bibinfo {volume}
  {13}},\ \bibinfo {pages} {085006} (\bibinfo {year} {2011})}\BibitemShut
  {NoStop}%
\bibitem [{\citenamefont {Hodgman}\ \emph {et~al.}(2009)\citenamefont
  {Hodgman}, \citenamefont {Dall}, \citenamefont {Byron}, \citenamefont
  {Baldwin}, \citenamefont {Buckman},\ and\ \citenamefont
  {Truscott}}]{PhysRevLett.103.053002}%
  \BibitemOpen
  \bibfield  {author} {\bibinfo {author} {\bibfnamefont {S.~S.}\ \bibnamefont
  {Hodgman}}, \bibinfo {author} {\bibfnamefont {R.~G.}\ \bibnamefont {Dall}},
  \bibinfo {author} {\bibfnamefont {L.~J.}\ \bibnamefont {Byron}}, \bibinfo
  {author} {\bibfnamefont {K.~G.~H.}\ \bibnamefont {Baldwin}}, \bibinfo
  {author} {\bibfnamefont {S.~J.}\ \bibnamefont {Buckman}},\ and\ \bibinfo
  {author} {\bibfnamefont {A.~G.}\ \bibnamefont {Truscott}},\ }\bibfield
  {title} {\bibinfo {title} {Metastable helium: A new determination of the
  longest atomic excited-state lifetime},\ }\href
  {https://doi.org/10.1103/PhysRevLett.103.053002} {\bibfield  {journal}
  {\bibinfo  {journal} {Phys. Rev. Lett.}\ }\textbf {\bibinfo {volume} {103}},\
  \bibinfo {pages} {053002} (\bibinfo {year} {2009})}\BibitemShut {NoStop}%
\bibitem [{\citenamefont {Dall}\ and\ \citenamefont
  {Truscott}(2007)}]{DALL2007255}%
  \BibitemOpen
  \bibfield  {author} {\bibinfo {author} {\bibfnamefont {R.}~\bibnamefont
  {Dall}}\ and\ \bibinfo {author} {\bibfnamefont {A.}~\bibnamefont
  {Truscott}},\ }\bibfield  {title} {\bibinfo {title} {Bose-einstein
  condensation of metastable helium in a bi-planar quadrupole ioffe
  configuration trap},\ }\href
  {https://doi.org/https://doi.org/10.1016/j.optcom.2006.09.031} {\bibfield
  {journal} {\bibinfo  {journal} {Optics Communications}\ }\textbf {\bibinfo
  {volume} {270}},\ \bibinfo {pages} {255 } (\bibinfo {year}
  {2007})}\BibitemShut {NoStop}%
\bibitem [{\citenamefont {Lin}(2004)}]{ChunLinphd}%
  \BibitemOpen
  \bibfield  {author} {\bibinfo {author} {\bibfnamefont {E.~C.}\ \bibnamefont
  {Lin}},\ }\emph {\bibinfo {title} {CompressedSensing for Electronic Radio
  Frequency Receiver:Detection, Sensitivity, and Implementation}},\ \href
  {https://corescholar.libraries.wright.edu/cgi/viewcontent.cgi?referer=https://www.google.com/&httpsredir=1&article=2603&context=etd_all}
  {Ph.D. thesis},\ \bibinfo  {school} {Yuan-Ze University} (\bibinfo {year}
  {2004})\BibitemShut {NoStop}%
\bibitem [{\citenamefont {{Draganic}}\ \emph {et~al.}(2017)\citenamefont
  {{Draganic}}, \citenamefont {{Orovic}},\ and\ \citenamefont
  {{Stankovic}}}]{2017arXiv170505216D}%
  \BibitemOpen
  \bibfield  {author} {\bibinfo {author} {\bibfnamefont {A.}~\bibnamefont
  {{Draganic}}}, \bibinfo {author} {\bibfnamefont {I.}~\bibnamefont
  {{Orovic}}},\ and\ \bibinfo {author} {\bibfnamefont {S.}~\bibnamefont
  {{Stankovic}}},\ }\bibfield  {title} {\bibinfo {title} {{On some common
  compressive sensing recovery algorithms and applications - Review paper}},\
  }\href@noop {} {\bibfield  {journal} {\bibinfo  {journal} {arXiv e-prints}\
  ,\ \bibinfo {eid} {arXiv:1705.05216}} (\bibinfo {year} {2017})},\ \Eprint
  {https://arxiv.org/abs/1705.05216} {arXiv:1705.05216 [cs.IT]} \BibitemShut
  {NoStop}%
\bibitem [{\citenamefont {{Kallinger, T.}}\ \emph {et~al.}(2008)\citenamefont
  {{Kallinger, T.}}, \citenamefont {{Reegen, P.}},\ and\ \citenamefont {{Weiss,
  W. W.}}}]{refld0}%
  \BibitemOpen
  \bibfield  {author} {\bibinfo {author} {\bibnamefont {{Kallinger, T.}}},
  \bibinfo {author} {\bibnamefont {{Reegen, P.}}},\ and\ \bibinfo {author}
  {\bibnamefont {{Weiss, W. W.}}},\ }\bibfield  {title} {\bibinfo {title} {A
  heuristic derivation of the uncertainty for frequency determination in time
  series data},\ }\href {https://doi.org/10.1051/0004-6361:20077559} {\bibfield
   {journal} {\bibinfo  {journal} {A\&A}\ }\textbf {\bibinfo {volume} {481}},\
  \bibinfo {pages} {571} (\bibinfo {year} {2008})}\BibitemShut {NoStop}%
\bibitem [{\citenamefont {Dedman}\ \emph {et~al.}(2007)\citenamefont {Dedman},
  \citenamefont {Dall}, \citenamefont {Byron},\ and\ \citenamefont
  {Truscott}}]{Dedman2007}%
  \BibitemOpen
  \bibfield  {author} {\bibinfo {author} {\bibfnamefont {C.~J.}\ \bibnamefont
  {Dedman}}, \bibinfo {author} {\bibfnamefont {R.~G.}\ \bibnamefont {Dall}},
  \bibinfo {author} {\bibfnamefont {L.~J.}\ \bibnamefont {Byron}},\ and\
  \bibinfo {author} {\bibfnamefont {A.~G.}\ \bibnamefont {Truscott}},\
  }\bibfield  {title} {\bibinfo {title} {Active cancellation of stray magnetic
  fields in a bose-einstein condensation experiment},\ }\href
  {https://doi.org/http://dx.doi.org/10.1063/1.2472600} {\bibfield  {journal}
  {\bibinfo  {journal} {Rev Sci Instrum}\ }\textbf {\bibinfo {volume} {78}},\
  \bibinfo {eid} {024703} (\bibinfo {year} {2007})}\BibitemShut {NoStop}%
\bibitem [{\citenamefont {Henson}\ \emph
  {et~al.}(2018{\natexlab{b}})\citenamefont {Henson}, \citenamefont {Yue},
  \citenamefont {Hodgman}, \citenamefont {Shin}, \citenamefont {Smirnov},
  \citenamefont {Ostrovskaya}, \citenamefont {Guan},\ and\ \citenamefont
  {Truscott}}]{PhysRevA.97.063601}%
  \BibitemOpen
  \bibfield  {author} {\bibinfo {author} {\bibfnamefont {B.~M.}\ \bibnamefont
  {Henson}}, \bibinfo {author} {\bibfnamefont {X.}~\bibnamefont {Yue}},
  \bibinfo {author} {\bibfnamefont {S.~S.}\ \bibnamefont {Hodgman}}, \bibinfo
  {author} {\bibfnamefont {D.~K.}\ \bibnamefont {Shin}}, \bibinfo {author}
  {\bibfnamefont {L.~A.}\ \bibnamefont {Smirnov}}, \bibinfo {author}
  {\bibfnamefont {E.~A.}\ \bibnamefont {Ostrovskaya}}, \bibinfo {author}
  {\bibfnamefont {X.~W.}\ \bibnamefont {Guan}},\ and\ \bibinfo {author}
  {\bibfnamefont {A.~G.}\ \bibnamefont {Truscott}},\ }\bibfield  {title}
  {\bibinfo {title} {Bogoliubov-cherenkov radiation in an atom laser},\ }\href
  {https://doi.org/10.1103/PhysRevA.97.063601} {\bibfield  {journal} {\bibinfo
  {journal} {Phys. Rev. A}\ }\textbf {\bibinfo {volume} {97}},\ \bibinfo
  {pages} {063601} (\bibinfo {year} {2018}{\natexlab{b}})}\BibitemShut
  {NoStop}%
\bibitem [{\citenamefont {Dedman}(2016)}]{colincurrentctr}%
  \BibitemOpen
  \bibfield  {author} {\bibinfo {author} {\bibfnamefont {C.}~\bibnamefont
  {Dedman}},\ }\bibfield  {title} {\bibinfo {title} {High-stability, low-noise
  current driver},\ }\href {http://www.vishaypg.com/docs/63627/63627.pdf}
  {\bibfield  {journal} {\bibinfo  {journal} {Vishay Case Studies}\ } (\bibinfo
  {year} {2016})}\BibitemShut {NoStop}%
\bibitem [{\citenamefont {Abbas}\ \emph {et~al.}(2021)\citenamefont {Abbas},
  \citenamefont {Meng}, \citenamefont {Patil}, \citenamefont {Ross},
  \citenamefont {Truscott},\ and\ \citenamefont
  {Hodgman}}]{PhysRevA.103.053317}%
  \BibitemOpen
  \bibfield  {author} {\bibinfo {author} {\bibfnamefont {A.~H.}\ \bibnamefont
  {Abbas}}, \bibinfo {author} {\bibfnamefont {X.}~\bibnamefont {Meng}},
  \bibinfo {author} {\bibfnamefont {R.~S.}\ \bibnamefont {Patil}}, \bibinfo
  {author} {\bibfnamefont {J.~A.}\ \bibnamefont {Ross}}, \bibinfo {author}
  {\bibfnamefont {A.~G.}\ \bibnamefont {Truscott}},\ and\ \bibinfo {author}
  {\bibfnamefont {S.~S.}\ \bibnamefont {Hodgman}},\ }\bibfield  {title}
  {\bibinfo {title} {Rapid generation of metastable helium {B}ose-{E}instein
  condensates},\ }\href {https://doi.org/10.1103/PhysRevA.103.053317}
  {\bibfield  {journal} {\bibinfo  {journal} {Phys. Rev. A}\ }\textbf {\bibinfo
  {volume} {103}},\ \bibinfo {pages} {053317} (\bibinfo {year}
  {2021})}\BibitemShut {NoStop}%
\bibitem [{\citenamefont {Debs}\ \emph {et~al.}(2009)\citenamefont {Debs},
  \citenamefont {D\"{o}ring}, \citenamefont {Robins}, \citenamefont {Figl},
  \citenamefont {Altin},\ and\ \citenamefont {Close}}]{Debs2009}%
  \BibitemOpen
  \bibfield  {author} {\bibinfo {author} {\bibfnamefont {J.~E.}\ \bibnamefont
  {Debs}}, \bibinfo {author} {\bibfnamefont {D.}~\bibnamefont {D\"{o}ring}},
  \bibinfo {author} {\bibfnamefont {N.~P.}\ \bibnamefont {Robins}}, \bibinfo
  {author} {\bibfnamefont {C.}~\bibnamefont {Figl}}, \bibinfo {author}
  {\bibfnamefont {P.~A.}\ \bibnamefont {Altin}},\ and\ \bibinfo {author}
  {\bibfnamefont {J.~D.}\ \bibnamefont {Close}},\ }\bibfield  {title} {\bibinfo
  {title} {A two-state raman coupler for coherent atom optics},\ }\href
  {https://doi.org/10.1364/oe.17.002319} {\bibfield  {journal} {\bibinfo
  {journal} {Optics Express}\ }\textbf {\bibinfo {volume} {17}},\ \bibinfo
  {pages} {2319} (\bibinfo {year} {2009})}\BibitemShut {NoStop}%
\bibitem [{\citenamefont {Bücker}\ \emph {et~al.}(2009)\citenamefont
  {Bücker}, \citenamefont {Perrin}, \citenamefont {Manz}, \citenamefont
  {Betz}, \citenamefont {Koller}, \citenamefont {Plisson}, \citenamefont
  {Rottmann}, \citenamefont {Schumm},\ and\ \citenamefont
  {Schmiedmayer}}]{lightsheetSchmidemayer}%
  \BibitemOpen
  \bibfield  {author} {\bibinfo {author} {\bibfnamefont {R.}~\bibnamefont
  {Bücker}}, \bibinfo {author} {\bibfnamefont {A.}~\bibnamefont {Perrin}},
  \bibinfo {author} {\bibfnamefont {S.}~\bibnamefont {Manz}}, \bibinfo {author}
  {\bibfnamefont {T.}~\bibnamefont {Betz}}, \bibinfo {author} {\bibfnamefont
  {C.}~\bibnamefont {Koller}}, \bibinfo {author} {\bibfnamefont
  {T.}~\bibnamefont {Plisson}}, \bibinfo {author} {\bibfnamefont
  {J.}~\bibnamefont {Rottmann}}, \bibinfo {author} {\bibfnamefont
  {T.}~\bibnamefont {Schumm}},\ and\ \bibinfo {author} {\bibfnamefont
  {J.}~\bibnamefont {Schmiedmayer}},\ }\bibfield  {title} {\bibinfo {title}
  {Single-particle-sensitive imaging of freely propagating ultracold atoms},\
  }\href {https://doi.org/10.1088/1367-2630/11/10/103039} {\bibfield  {journal}
  {\bibinfo  {journal} {New Journal of Physics}\ }\textbf {\bibinfo {volume}
  {11}},\ \bibinfo {pages} {103039} (\bibinfo {year} {2009})}\BibitemShut
  {NoStop}%
\bibitem [{\citenamefont {Fedichev}\ \emph {et~al.}(1998)\citenamefont
  {Fedichev}, \citenamefont {Shlyapnikov},\ and\ \citenamefont
  {Walraven}}]{PhysRevLett.80.2269}%
  \BibitemOpen
  \bibfield  {author} {\bibinfo {author} {\bibfnamefont {P.~O.}\ \bibnamefont
  {Fedichev}}, \bibinfo {author} {\bibfnamefont {G.~V.}\ \bibnamefont
  {Shlyapnikov}},\ and\ \bibinfo {author} {\bibfnamefont {J.~T.~M.}\
  \bibnamefont {Walraven}},\ }\bibfield  {title} {\bibinfo {title} {Damping of
  low-energy excitations of a trapped {B}ose-{E}instein condensate at finite
  temperatures},\ }\href {https://doi.org/10.1103/PhysRevLett.80.2269}
  {\bibfield  {journal} {\bibinfo  {journal} {Phys. Rev. Lett.}\ }\textbf
  {\bibinfo {volume} {80}},\ \bibinfo {pages} {2269} (\bibinfo {year}
  {1998})}\BibitemShut {NoStop}%
\bibitem [{\citenamefont {Yuen}\ \emph {et~al.}(2015)\citenamefont {Yuen},
  \citenamefont {Barr}, \citenamefont {Cotter}, \citenamefont {Butler},\ and\
  \citenamefont {Hinds}}]{Yuen_2015}%
  \BibitemOpen
  \bibfield  {author} {\bibinfo {author} {\bibfnamefont {B.}~\bibnamefont
  {Yuen}}, \bibinfo {author} {\bibfnamefont {I.~J.~M.}\ \bibnamefont {Barr}},
  \bibinfo {author} {\bibfnamefont {J.~P.}\ \bibnamefont {Cotter}}, \bibinfo
  {author} {\bibfnamefont {E.}~\bibnamefont {Butler}},\ and\ \bibinfo {author}
  {\bibfnamefont {E.~A.}\ \bibnamefont {Hinds}},\ }\bibfield  {title} {\bibinfo
  {title} {Enhanced oscillation lifetime of a {B}ose{\textendash}{E}instein
  condensate in the 3d/1d crossover},\ }\href
  {https://doi.org/10.1088/1367-2630/17/9/093041} {\bibfield  {journal}
  {\bibinfo  {journal} {New Journal of Physics}\ }\textbf {\bibinfo {volume}
  {17}},\ \bibinfo {pages} {093041} (\bibinfo {year} {2015})}\BibitemShut
  {NoStop}%
\bibitem [{\citenamefont {Weisstein}(2021)}]{HarmonicAdd}%
  \BibitemOpen
  \bibfield  {author} {\bibinfo {author} {\bibfnamefont {E.}~\bibnamefont
  {Weisstein}},\ }\bibfield  {title} {\bibinfo {title} {Harmonic addition
  theorem},\ }\href
  {https://mathworld.wolfram.com/HarmonicAdditionTheorem.html} {\bibfield
  {journal} {\bibinfo  {journal} {MathWorld--A Wolfram Web Resource}\ }
  (\bibinfo {year} {2021})}\BibitemShut {NoStop}%
\bibitem [{\citenamefont {Pethick}\ and\ \citenamefont
  {Smith}(2001)}]{Pethick}%
  \BibitemOpen
  \bibfield  {author} {\bibinfo {author} {\bibfnamefont {C.~J.}\ \bibnamefont
  {Pethick}}\ and\ \bibinfo {author} {\bibfnamefont {H.}~\bibnamefont
  {Smith}},\ }\href {https://doi.org/10.1017/CBO9780511755583} {\emph {\bibinfo
  {title} {Bose–Einstein Condensation in Dilute Gases}}}\ (\bibinfo
  {publisher} {Cambridge University Press},\ \bibinfo {year}
  {2001})\BibitemShut {NoStop}%
\bibitem [{\citenamefont {Moal}\ \emph {et~al.}(2006)\citenamefont {Moal},
  \citenamefont {Portier}, \citenamefont {Kim}, \citenamefont {Dugu\'e},
  \citenamefont {Rapol}, \citenamefont {Leduc},\ and\ \citenamefont
  {Cohen-Tannoudji}}]{PhysRevLett.96.023203}%
  \BibitemOpen
  \bibfield  {author} {\bibinfo {author} {\bibfnamefont {S.}~\bibnamefont
  {Moal}}, \bibinfo {author} {\bibfnamefont {M.}~\bibnamefont {Portier}},
  \bibinfo {author} {\bibfnamefont {J.}~\bibnamefont {Kim}}, \bibinfo {author}
  {\bibfnamefont {J.}~\bibnamefont {Dugu\'e}}, \bibinfo {author} {\bibfnamefont
  {U.~D.}\ \bibnamefont {Rapol}}, \bibinfo {author} {\bibfnamefont
  {M.}~\bibnamefont {Leduc}},\ and\ \bibinfo {author} {\bibfnamefont
  {C.}~\bibnamefont {Cohen-Tannoudji}},\ }\bibfield  {title} {\bibinfo {title}
  {Accurate determination of the scattering length of metastable helium atoms
  using dark resonances between atoms and exotic molecules},\ }\href
  {https://doi.org/10.1103/PhysRevLett.96.023203} {\bibfield  {journal}
  {\bibinfo  {journal} {Phys. Rev. Lett.}\ }\textbf {\bibinfo {volume} {96}},\
  \bibinfo {pages} {023203} (\bibinfo {year} {2006})}\BibitemShut {NoStop}%
\bibitem [{\citenamefont {{Montgomery}}\ and\ \citenamefont
  {{Odonoghue}}(1999)}]{Montgomery1999}%
  \BibitemOpen
  \bibfield  {author} {\bibinfo {author} {\bibfnamefont {M.~H.}\ \bibnamefont
  {{Montgomery}}}\ and\ \bibinfo {author} {\bibfnamefont {D.}~\bibnamefont
  {{Odonoghue}}},\ }\bibfield  {title} {\bibinfo {title} {{A derivation of the
  errors for least squares fitting to time series data}},\ }\href@noop {}
  {\bibfield  {journal} {\bibinfo  {journal} {Delta Scuti Star Newsletter}\
  }\textbf {\bibinfo {volume} {13}},\ \bibinfo {pages} {28} (\bibinfo {year}
  {1999})}\BibitemShut {NoStop}%
\bibitem [{Note1()}]{Note1}%
  \BibitemOpen
  \bibinfo {note} {The ratio of widths for a pulse removed from a BEC which has
  an atom number 10\% of the starting value is a factor of 0.63 smaller. For a
  series of PALs where the atom number linearly decreases to a final atom
  number 10\% of the starting value, the average spatial width of the pulses
  will be 0.87 that of the first.}\BibitemShut {Stop}%
\end{thebibliography}%

\end{document}